\documentclass[12pt,hidelinks]{article}
\usepackage[body={17.5cm, 22cm},right=2cm]{geometry}
\usepackage{color}
\usepackage{amssymb,graphicx}
\usepackage{amsmath}
\usepackage{epsfig}
\usepackage[bookmarks=true,pdfborder={0 0 0}]{hyperref} 
\usepackage{cite}

\usepackage{mathtools,empheq,slashed,braket,array}
\usepackage{setspace}
\usepackage{subfigure}
\usepackage{hyperref}
\usepackage{tabularx}
\usepackage[makeroom]{cancel}

\newcommand{\eV}{{\, {\rm eV}}}
\newcommand{\keV}{{\, {\rm keV}}}
\newcommand{\MeV}{{\, {\rm MeV}}}

\usepackage{float}
\allowdisplaybreaks

\newcommand{\cL}{\mathcal{L}}

\newcommand{\E}{\textbf{E}}
\newcommand{\B}{\textbf{B}}

\newcommand{\eg}{\textit{e.g.}}
\newcommand{\ie}{\textit{i.e.}}

\newcommand\cmt[1]{}
\def\shrug{\texttt{\raisebox{0.75em}{\char`\_}\char`\\\char`\_\kern-0.5ex(\kern-0.25ex\raisebox{0.25ex}{\rotatebox{45}{\raisebox{-.75ex}"\kern-1.5ex\rotatebox{-90})}}\kern-0.5ex)\kern-0.5ex\char`\_/\raisebox{0.75em}{\char`\_}}}
\usepackage{accents}

\usepackage{physics}  
\usepackage[compat=1.1.0]{tikz-feynman}
\usepackage{environ}      

\newcommand{\leri}[1]{\left(#1 \right)}

\usepackage[nolist]{acronym}
\begin{acronym}
	\acro{SM}{Standard Model}
	\acro{CPV}{CP violating}
	\acro{CPC}{CP conserving}
	\acro{EH}{Euler-Heisenberg}
	\acro{BSM}{beyond the Standard Model}
	\acro{COM}{center of mass}
	\acro{QED}{Quantum Electrodynamics}
	\acro{EFT}{Effective Field Theory}
	\acro{ALP}{axion-like particle}
	\acro{EOM}{equation of motion}
	\acro{RHS}{right hand side}
	\acro{LHS}{left hand side}
	\acro{FP}{Fabry-Perot}
	\acro{EM}{Electromagnetism}
	\acro{SRF}{superconducting radio frequency}
	\acro{SNR}{signal to noise ratio}
	\acro{EP}{Equivalence Principle}
	\acro{CL}{confidence level}
\end{acronym}

\newcounter{mysub}
\renewcommand{\themysub}{\roman{mysub}}

\begin{document}
\setcounter{page}{0}
\thispagestyle{empty}

\parskip 3pt

\font\mini=cmr10 at 2pt

\begin{titlepage}
~\vspace{1cm}
\begin{center}

{\LARGE \bf Probing CP Violation in Photon \\
\vspace{2.6mm}
 Self-Interactions with Cavities}

\vspace{0.6cm}

{\large
Marco~Gorghetto$^a$,
Gilad~Perez$^a$,  
Inbar~Savoray$^a$,  
and Yotam~Soreq$^{b}$}
\\
\vspace{.6cm}
{\normalsize { \sl $^{a}$ 
Department of Particle Physics and Astrophysics, Weizmann Institute of Science,\\ Rehovot 761001, Israel }}

\vspace{.3cm}
{\normalsize { \sl $^{b}$ Physics Department, Technion -- Israel Institute of Technology, Haifa 3200003, Israel}}

\end{center}
\vspace{.8cm}

\begin{abstract}
In this paper we study CP violation in photon self-interactions at low energy. 
These interactions, mediated by the effective operator $FFF\tilde{F}$, where ($\tilde F$) $F$ is the (dual) electromagnetic field strength, have yet to be directly probed experimentally. 
Possible sources for such interactions are weakly coupled light scalars with both scalar and pseudoscalar couplings to photons (for instance, complex Higgs-portal scalars or the relaxion), or new light fermions coupled to photons via dipole operators. 
We propose a method to isolate the CP-violating contribution to the photon self-interactions using Superconducting Radio-Frequency cavities and vacuum birefringence experiments. In addition, we consider several theoretical and experimental indirect bounds on the scale of new physics associated with the above effective operator, and present projections for the sensitivity of the proposed experiments to this scale. We also discuss the implications of these bounds on the CP-violating couplings of new light particles coupled to photons.

\end{abstract}

\end{titlepage}

{\fontsize{11}{10}
\tableofcontents
}


\section{Introduction} 
\label{sec:intro}

Photon self-interactions are absent in pure Maxwell's theory, but are induced by photon-matter interactions. Therefore, effective photon-photon interactions appear in generic low energy \ac{QED} theories. 
The \ac{EH} Lagrangian is an example of such an effective theory, where the non-linear term arises at one loop, at energies below the electron mass~\cite{Heisenberg:1935qt,Schwinger:1951nm}, and preserves CP. 
\ac{CPV} sources of photon interactions, on the other hand, arise in the \ac{SM} only at multiple loop level from the, CKM-suppressed, weak interactions, or are controlled by the tiny strong CP phase, and are thus negligibly small~\cite{Millo:2008ug}. 
Given the suppression of the \ac{SM} contribution, it should in principle be possible to probe (and possibly discover) \ac{CPV} new physics in this channel, without a significant \ac{SM} background.

As suggested in~\cite{Sikivie:1983ip}, the QCD axion, or in general \acp{ALP}, are well motivated new sources of photon self-interactions. 
Depending on their mass and their coupling to photons, the contributions of such particles to effective photon-photon interactions -- which are \ac{CPC} -- may be comparable to that of the \ac{EH} term or even exceed it~\cite{Evans:2018qwy,Bernard:1997kj}. 
\ac{CPV} photon self-interactions are instead mediated by degrees of freedom that are not CP-eigenstates. 
Scalars of this kind appear in theoretically motivated models such as the complex Higgs portal~\cite{Piazza:2010ye} and the relaxion~\cite{Graham:2015cka,Flacke:2016szy}. 
These interactions could also be induced by new fermions with non-vanishing electric and magnetic dipole moments. 
In the following we will mostly adopt a model-independent approach, where the information of the particular new physics providing photon self-interactions will be encoded in the the Wilson coefficients of the effective photon operators.

Effective \ac{CPC} and \ac{CPV} photon self-interactions \ac{BSM} can be indirectly constrained by the measured electronic magnetic dipole moment~\cite{Hanneke:2008tm} (see~\cite{Aoyama:2014sxa} for the \ac{SM} theory calculation), and by the upper limit on the electronic electric dipole moment~\cite{Andreev:2018ayy}. As we will show, the corresponding bounds can be easily estimated at energies above the electron mass, and yield relatively strong constraints on the scale of new physics. In contrast, at energies below the electron mass, the reach of direct experimental tests for the presence non-linear photon dynamics is, to date, much more limited. 
In fact, as reviewed in Section~\ref{sec:boundsFF}, current experiments, looking for vacuum birefringence (\eg~\cite{Ejlli:2020yhk}) and the Lamb shift~\cite{Wichmann:1956zz,Mohr:2012tt}, have probed only the \ac{CPC} part of the photon self-couplings. 
While their sensitivity is still about one order of magnitude above the \ac{EH} contribution, other experimental proposals such as~\cite{Eriksson:2004cz,Bogorad:2019pbu,Gao:2020anb,Salnikov:2020urr} could be able to measure such a term, and possibly constrain new physics contribution to light-by-light interactions.

In this paper we study the prospects of directly detecting \ac{CPV} photon self-interactions at energies below the electron mass, described by the effective operator $F_{\mu\nu}F^{\mu\nu}F_{\rho\sigma}\tilde{F}^{\rho\sigma}$, which has not been directly probed by any current experiment. 
In particular, we present simple modifications to proposed and currently running experimental setups, such that they can be made sensitive also to \ac{CPV} phenomena. 
Crucially, our proposals will be able to disentangle the \ac{CPV} contribution from the \ac{CPC} one, providing unique probes of the \ac{CPV} operator that, as mentioned, are free from the \ac{SM} background. These experiments could set the first model-independent bound on \ac{CPV} effective photon interactions at energies below the electron mass.

Our first proposal employs the production and detection of light-by-light interactions in a \ac{SRF} cavity, extending the setup described in~\cite{Brodin:2001zz,Eriksson:2004cz,Bogorad:2019pbu}. 
In this configuration, the self-interactions of background resonance modes pumped into the cavity act as a source, exciting another resonance mode of the cavity. 
We demonstrate how a particular choice of the pump and signal modes and of the cavity geometry allows singling out \ac{CPV} photon self-interactions. 

Photon nonlinearities are known to induce vacuum birefringence in the presence of an external electromagnetic field. 
In particular, polarized light acquires a non-vanishing ellipticity and rotation of the polarization plane~\cite{PhysRevD.75.117301,Liao:2007nu}. 
As a second probe, we discuss an experimental configuration where this phenomenon happens in a ring cavity. 
While inspired by the linear \ac{FP} cavity of the PVLAS experiment~\cite{DellaValle:2015xxa}, which is essentially insensitive to CP-odd effects, we show that a ring cavity geometry is sensitive to \ac{CPC} and \ac{CPV} photon interactions simultaneously, which can be distinguished by a temporal analysis of the signal. A similar scheme has been proposed in~\cite{Fan:2017sxk}, and applied to  \ac{CPV} dark sectors.

The paper is organized as follows: 
In Section~\ref{sec:boundsFF} we define the photon \ac{EFT} at low energy and summarize the current direct and indirect bounds on its coefficients. 
We also discuss the possible contributions to photon self-interactions from simple new physics models. 
In Section~\ref{sec:SRF} we discuss the prospects of detection of \ac{CPV} photon interactions using an \ac{SRF} cavity. 
In Section~\ref{sec:RingCavity} we study the detection of vacuum birefringence and dichroism in a ring cavity, and its implication for the \ac{CPV} operator. 
We conclude in Section~\ref{sec:conclusions}.

\section[Photon \acl{EFT}  and Current Bounds]{Photon \ac{EFT}  and Current Bounds}
\label{sec:boundsFF}

\subsection[Agnostic \acl{EFT} Approach]{Agnostic \ac{EFT} Approach}

At energies below the electron mass, interactions among photons are self-consistently described by an effective Lagrangian involving the photon field only. 
Such a Lagrangian can be conveniently expanded in powers of the two independent gauge invariant CP-even and CP-odd operators, $\frac12 F_{\mu\nu}F^{\mu\nu}=\textbf{E}^2-\textbf{B}^2$ and $\frac14 F_{\mu\nu}\tilde{F}^{\mu\nu}=\textbf{E}\cdot\textbf{B}$, where $F_{\mu\nu}$ and $\tilde{F}_{\mu\nu}$ are the photon field strength and its dual, and $\E\,(\B)$ is the electric\,(magnetic) field. 
At leading order, up to dimension-8, it reads~(see \eg~\cite{Millo:2008ug})
\begin{align}
	\label{eq:LEFT}
	\cL_{\rm EFT}
=	-\frac14F_{\mu\nu}F^{\mu\nu}+\frac{a}{4} F_{\mu\nu}\tilde{F}^{\mu\nu}
	+\frac{b}{4}(F_{\mu\nu}F^{\mu\nu})^2
	+\frac{c}{16}(F_{\mu\nu}\tilde{F}^{\mu\nu})^2
	+\frac{d}{8}F_{\mu\nu}F^{\mu\nu}F_{\rho\sigma}\tilde{F}^{\rho\sigma} \ .
\end{align}
Note that $a$ has no physical effect being $F_{\mu\nu}\tilde{F}^{\mu\nu}$ a total derivative. 
The coefficients $b,c,d$ are proportional to four inverse powers of the \ac{EFT} UV-cutoff, $\Lambda$. 

In the \ac{SM} the coefficients $b$ and $c$ receive leading contribution from the \ac{EH} effective action~\cite{Heisenberg:1935qt,Schwinger:1951nm}
\begin{align}
	\label{eq:bcEH}
	b_{\rm EH}
=	\frac{2}{45}\frac{\alpha_{\rm EM}^2}{m_e^4}\approx (13\,\MeV)^{-4}\, ,  
	\qquad 
	c_{\rm EH}
=	7\,b_{\rm EH} \,,
\end{align}
where $m_e$ is the electron mass and $\alpha_{\rm EM}$ is the fine-structure constant. The term 
$ F_{\mu\nu}F^{\mu\nu}F_{\rho\sigma}\tilde{F}^{\rho\sigma}$ violates CP and obtains radiative contributions from the two \ac{CPV} sources of the \ac{SM}. 
First, the contribution from the QCD $\theta$-term has been estimated in chiral perturbation theory in the large $N_c$ limit in Ref.~\cite{Millo:2008ug}. 
Due to the smallness of $\theta\lesssim10^{-10}$, it is suppressed by at least $20$ orders of magnitudes with respect to the \ac{EH} \ac{CPC} self-interactions in Eq.~\eqref{eq:bcEH}. 
A second contribution to $d$ comes from the \ac{CPV} phase of the CKM matrix. 
Although never calculated, to the best of our knowledge it is expected to appear at least at three loops and (due to the GIM mechanism) be extremely suppressed. 
Given Eq.~\eqref{eq:bcEH} and the smallness of $d$, the UV cutoff $\Lambda$
is therefore of order $10\,\MeV$. 

As discussed in the next section, the coefficients of the Lagrangian in Eq.~\eqref{eq:LEFT} could obtain contributions from \ac{BSM} physics. 
On theoretical grounds, they are subject to positivity constraints if $\cL_{\rm EFT}$ comes from a causal UV theory with an analytic and unitary $S$-matrix. 
In particular, $b$ and $c$ must be positive, and $d$ must be bounded~\cite{Remmen:2019cyz}
\begin{align}
	\label{eq:EFTcon}
	|d| < 2\sqrt{bc}\, .
\end{align}
This provides a nontrivial consistency condition on $\cL_{\rm EFT}$, that requires the magnitude of the CP-odd term to be bounded by the CP-even ones. 
In particular, if a violation of this bound is measured, one would need to give up some of the fundamental principles in the UV theory underlying the effective Lagrangian in Eq.~\eqref{eq:LEFT}, such as analyticity, unitarity or causality.

Let us now review the current experimental limits on $\cL_{\rm EFT}$ at energies below the \ac{EH} scale.\footnote{Within the \ac{EFT} in Eq.~\eqref{eq:EFTcon} the Wilson coefficients $b,c,d$ depend on the energy scale and are expected to be subject to logarithmic running and mixing effects. Since the running is only logarithmic, we will tacitly ignore this dependence. When discussing bounds, we will imply that the coefficients are evaluated at the energy scale corresponding to the typical frequency of the experiment.}
The coefficients $b$ and $c$ are hardly constrained from direct light-by-light scattering~\cite{Bernard:2010dx,Fouche:2016qqj}, which provides limits that are more than $10$ orders of magnitude above the \ac{EH} prediction in Eq.~\eqref{eq:bcEH}. 
The coefficient $b$ induces a correction to the Coulomb potential of the hydrogen atom~\cite{Wichmann:1956zz} and a Lamb shift of its 1S and 2S energy levels. 
For the \ac{EH} Lagrangian, this correction to the energy levels has been calculated to be $3\times10^{-4}$ times the leading term~\cite{Mohr:2012tt}, while the related measurements still have a precision of $3\times10^{-3}$~\cite{Biraben:2001uqg}. 
Since the correction is linear with $b$, this therefore yields a bound of 
\begin{align}
	\label{eq:bLambShift}
	\abs{b}\lesssim 20 \, b_{\rm EH} \, ,
\end{align}
at $95\%$ confidence level.

As explained in more detail in Section~\ref{sec:RingCavity}, the combination $c-4b$ induces a non-vanishing ellipticity to polarized light passing through a region permeated by an external magnetic field. 
This observable has been bounded by BMV~\cite{BMV} and PVLAS~\cite{DellaValle:2015xxa}, and the latest limit~\cite{Ejlli:2020yhk} corresponds to
\begin{equation}
	\label{eq:bcPVLAS}
	\frac{c-4b}{c_\text{EH}-4b_\text{EH}}=4.8\pm 6.8\, ,
\end{equation}
where $c_\text{EH}-4b_\text{EH}=3b_\text{EH}$. 
We obtain a bound on $c$ by combining the $2\,\sigma$ intervals of Eqs.~\eqref{eq:bLambShift} and~\eqref{eq:bcPVLAS}

\begin{equation}
	\label{eq:cComb}
	-15\,c_\text{EH}\lesssim c\lesssim 19 \,c_\text{EH}\, .
\end{equation}
Eqs.~\eqref{eq:bLambShift} and~\eqref{eq:cComb} correspond to the strongest direct bounds to date. 

Currently, there is no direct experimental constraint on the coefficient $d$, although there have been theoretical studies~\cite{Millo:2008ug,DaSouza:2006abc} of its possible detection using vacuum birefringence. 
On the other hand, from Eqs.~\eqref{eq:bLambShift} and~\eqref{eq:cComb} we expect that the \ac{EFT} is consistent according to Eq.~\eqref{eq:EFTcon} only if 
\begin{align}
	|d| \ \lesssim \ 40\sqrt{b_{\rm EH} \, c_{\rm EH}} \,,
\end{align}
which will be denoted as the \ac{EFT} consistency bound. 

At energies above the electron mass, the electron must be included in the \ac{EFT}, and the measurements of its electric and magnetic dipole moments put an indirect constrain on the contributions of new physics to photon self-interactions. Indeed, in the electron-photon \ac{EFT}, a single insertion of the 4-photon operators generates a two loop contribution to the magnetic and electric dipole operators
\begin{align}
	\label{eq:aede}
	\frac{a_e}{4m_e} e\overline{\psi}_e\sigma_{\mu\nu}\psi_e F^{\mu\nu} \, ,   
	\quad \quad 
	-\frac{i}{2} d_e \overline{\psi}_e\sigma_{\mu\nu} \gamma_5\psi_e F^{\mu\nu} \, ,
\end{align}
(where $\psi_e$ is the electron field and $e\equiv\sqrt{4\pi\alpha_{\rm EM}}$ is the electric charge) which could make the coefficients $a_e$ and $d_e$ deviate from their \ac{SM} predictions. 
Although never evaluated directly, a crude estimate for the deviations induced by new physics via such a diagram is
\begin{align}\label{eq:dadd}
	\frac{e}{4m_e}\Delta a_e
\simeq 	C_1\frac{e^3}{(16\pi^2)^2} m^3_e b_\text{BSM} \, ,
	\quad\quad
	\Delta d_e
\simeq 	C_2\frac{e^3}{(16\pi^2)^2} m^3_e d_\text{BSM} \, ,
\end{align}
where $b_\text{BSM}, c_\text{BSM}$ and $d_\text{BSM}$ are the \ac{BSM} contributions to the coefficients, and $C_1$ and $C_2$ are order one factors. Assuming $C_1$ and $C_2$ of order one, and considering the current bounds on $|a_{\rm exp}-a_e^{\rm SM}|/a_e^{\rm SM}\lesssim 10^{-9}$~\cite{Aoyama:2014sxa} and $|d_e|< 1.1\times 10^{-29}\, e\times\text{cm}$~\cite{Andreev:2018ayy}, we obtain a rough estimate for the bounds on the \ac{BSM} contribution to the \ac{EFT} coefficients as
\begin{align}\label{eq:bcdbsm}
	b_\text{BSM},c_\text{BSM}
	\lesssim
	10^{-2} \, b_{\rm EH}\, ,   
	\quad\quad 
	d_\text{BSM}
	\lesssim 
	10^{-8} \, b_{\rm EH}\,.
\end{align} 
These are indirect bounds on $b_\text{BSM}$, $c_\text{BSM}$ and $d_{\rm BSM}$ that strongly constrain possible new physics heavier than the electron. In the following, however, we will focus on constraining these operators at energies below the electron mass, as the existing bounds of this type are quite weak (see Eqs.~\eqref{eq:bLambShift} and~\eqref{eq:cComb}). In particular, our proposals will be able to set the first model-independent bound on the coefficient $d$ at energies below the electron mass.

In principle, it should be possible to translate the bounds from the electric and magnetic dipole moments in Eq.~\eqref{eq:bcdbsm} (valid at energies above $m_e$) into bounds on the low energy photon \ac{EFT} in eq.~\eqref{eq:LEFT}, so that they can be applied to new physics lighter than the electron. This however requires matching the photon \ac{EFT} with the photon-electron Lagrangian, and is beyond the scope of our work.

We finally note that light-by-light scattering has been observed at the LHC in Pb-Pb collisions by the ATLAS and CMS collaborations~\cite{Aad:2019ock,Sirunyan:2018fhl,dEnterria:2013zqi}. 
The diphoton invariant mass relevant to this measurement is above 6\,GeV, thus well beyond the scale at which the \ac{EFT} in Eq.~\eqref{eq:LEFT} is valid. 
At such energies the \ac{EFT} coefficients for photon self-interactions, including their \ac{CPV} part, have been constrained to be $b_\text{BSM}, c_\text{BSM}, d_\text{BSM}\lesssim 10^{-10}~\text{GeV}^{-4}$, see \cite{Akmansoy:2018xvd}.

\subsection{Contributions from New Physics}

As we now show, new particles coupled to photons generically contribute to the effective Lagrangian in Eq.~\eqref{eq:LEFT}, and possibly to the \ac{CPV} coefficient $d$.

As a first example, we consider a real scalar singlet under $U(1)_{\rm EM}$ with mass $m_\phi$. 
Even if not coupled to the photon at the renormalizable level, couplings to the photon field strength are present in the dimension-5 Lagrangian:
\begin{align}
	\label{eq:Lphi}
	\cL_{\phi}
	\supset
	\frac12(\partial_\mu\phi)^2-\frac12m_\phi^2\phi^2
	+\frac{\tilde{g}}{4}\phi F_{\mu\nu}\tilde{F}^{\mu\nu}
	+\frac{g}{4}\phi F_{\mu\nu}F^{\mu\nu} \, .
\end{align}
If both $g$ and $\tilde{g}$ are nonzero, $\phi$ has no definite transformation properties under CP, and $\cL_{\phi}$ breaks CP explicitly. 
When $\phi$ is integrated out, $\cL_{\phi}$ provides the contributions to $b,c,d$
\begin{align}
	\label{eq:bcdphi}
	b_\phi
=	\frac{g^2}{8m_\phi^2} \, , 
	\qquad  
	c_\phi
=	\frac{\tilde{g}^2}{2m_\phi^2}  \, , 
	\qquad 
	d_\phi
=	\frac{g\tilde{g}}{2m_\phi^2} \, ,
\end{align}
where $d_\phi$ is proportional to the \ac{CPV} combination $g\tilde{g}$. 
For \acp{ALP}, $g=0$ (to leading order) and $\tilde{g}$ provides a contribution to $c$. This contribution is larger than the \ac{EH} term if $\tilde{g}/m_\phi\gtrsim\alpha/m_e^2\approx 1/(10\,\MeV)^2$.
The couplings $g$ and $\tilde{g}$ arise simultaneously in models in which the scalar is not a CP eigenstate. In particular, in relaxion models~\cite{Graham:2015cka}, the scalar coupling $g$ is nonzero and determined by the relaxion's mixing with the Higgs~\cite{Flacke:2016szy}, while the pseudoscalar coupling $\tilde{g}$ is related to the shift-symmetric nature of the axion. 

We observe that for small enough $m_\phi$ the coupling $\tilde{g}$ is bounded by several probes, including laboratory experiments and astrophysical considerations, see~\cite{Graham:2015ouw,Irastorza:2018dyq,Choi:2020rgn } for reviews.
The coupling $g$ is stringently constrained from from fifth force and equivalence principle tests, see~\cite{Adelberger:2003zx,Lee:2020zjt,Hees:2018fpg} and references therein.
Nevertheless $d_\phi$ in Eq.~\eqref{eq:bcdphi} might easily dominate over the \ac{SM} contribution (this, for comparison, was estimated to be in its QCD part $d_\text{QCD}\lesssim 10^{-27} \MeV^{-4}$ where we used $\theta_{\rm QCD}\lesssim10^{-10}$~\cite{Millo:2008ug}). For scalars heavier than $m_e$, the couplings $g$ and $\tilde{g}$ are bounded by Eq.~\eqref{eq:bcdbsm}, see also~\cite{Marciano:2016yhf,DiLuzio:2020oah} for the full calculation.

Another possibility is to consider a new \ac{SM} gauge singlet fermion, $\psi$, which at dimension-5 level will be coupled to photons via dipole operators as
\begin{equation}
	\label{eq:Lpsi}
	\mathcal{L}_{\psi}
	\supset 
	\overline{\psi}i\gamma^\mu \partial_\mu \psi-m_\psi\overline{\psi}\psi
	+D\overline{\psi}\sigma_{\mu\nu}\psi F^{\mu\nu}+i\tilde{D}\overline{\psi}\sigma_{\mu\nu}\gamma_5\psi F^{\mu\nu} \, ,
\end{equation}
where $D$ and $\tilde{D}$ are its magnetic and electric dipole moments respectively.\footnote{Charged \ac{CPV} fermions have been discussed in detail \eg~in \cite{Yamashita:2017scc}.} The coefficients $b,c,d$ receive threshold corrections from one loop diagrams with four insertions of $D$ and/or $\tilde{D}$. 
In particular, $ b_\psi\sim4(D^4+\tilde{D}^4)/(16\pi^2)\,$, $c_\psi\sim 64 D^2\tilde{D}^2/(16\pi^2)$ and $d_\psi\sim 32(D^2+\tilde{D}^2)D\tilde{D}/(16\pi^2)$. 
As a result, the bounds in Eqs.~\eqref{eq:bLambShift} and~\eqref{eq:cComb} for $b$ and $c$ already imply an upper bound on the magnetic dipole moment, \emph{i.e.} $D,\tilde{D} \lesssim 10^{-7} \, \eV^{-1}$ for new singlet fermions with mass eV $\lesssim m_\psi\lesssim10\,\MeV$.\footnote{This mass range is required in order for the photon \ac{EFT}  to be applied. In particular, the lower bound comes from the typical energy scale associated to the experiments leading to the constrained quoted in Eqs.~\eqref{eq:bLambShift} and~\eqref{eq:cComb}.}
Dipole operators of light neutral fermions are in any case directly constrained by astrophysical observations (see \eg~\cite{Heger:2008er}), as the operators in Eq.~\eqref{eq:Lpsi} induce plasmon decay and therefore additional cooling of stars, implying $D,\tilde{D}\lesssim 4\times10^{-11} \mu_B\approx 10^{-18}\,\eV^{-1}$, where $\mu_B=e/(2m_e)$.\footnote{There is also a (weaker) bound, initially derived in~\cite{Barbieri:1988nh}, from supernova explosion.}
Contrary to the previous one, this bound applies only for masses $m_\psi$ up to few hundreds keV, corresponding to the core plasma temperature of the stars. Moreover, for $m_\psi\gtrsim m_e$ such operators are (indirectly) constrained by the electric and magnetic dipole moments of the electron in Eq.~\eqref{eq:aede}. For instance, a nonvanishing $\tilde{D}$ gives a contribution to $d_e$ at two loops, which can be roughly estimated as $\Delta d_e\simeq m_e m_\psi \tilde{D}^3e^2/(16\pi^2)^2$, and thus the previously mentioned bound on $d_e$ yields $\tilde{D}\lesssim 10^{-10} \eV^{-1} (m_\psi/\MeV)^{1/3}$.

\section[Isolating CP-Violation in an \acs{SRF} Cavity]{Isolating CP-Violation in an \ac{SRF} Cavity}
\label{sec:SRF}

The photon self-interactions in Eq.~\eqref{eq:LEFT} introduce nonlinearities in Maxwell's equations. 
These nonlinearities act as a source for an electromagnetic field in the presence of background electromagnetic waves. 
In this section, we discuss the production and the detection of this field in a \acf{SRF} 
 cavity, and point out how the contribution from \ac{CPV} photon interactions can be singled out by an appropriate choice of the background fields and the cavity dimensions.

We consider a free background field $A_p^{\mu}$ (satisfying $\partial_\mu F_p^{\mu\nu}=0$), and split the total field as $F_p^{\mu\nu}+F^{\mu\nu}$. 
At leading order in the photon self-couplings $b$, $c$ and $d$, the equations of motion of the Lagrangian in Eq.~\eqref{eq:LEFT} become a source-equation for the field $F^{\mu\nu}$, \ie
\begin{equation}\label{eq:sourceEq}
\partial_\mu F^{\mu\nu}=\left(\nabla\cdot\mathbf{E},\nabla\times\mathbf{B}-\partial_t\mathbf{E}\right)=J^\nu(A_p)  \, ,
\end{equation}
where the effective current $J^\mu$ is a function of the background field only, and reads
\begin{equation}
\begin{split}
	\label{eq:JEFT1}
	J^\mu
=	&\, F_p^{\mu\nu}\partial_{\nu}\leri{2b {F_p}_{\rho\sigma}{F_p}^{\rho\sigma}
	+\frac{d}{2}{F_p}_{\rho\sigma}\tilde{F_p}^{\rho\sigma}}
	+\tilde{F_p}^{\mu\nu}\partial_{\nu}\leri{\frac{c}{2}{F_p}_{\rho\sigma}\tilde{F_p}^{\rho\sigma}+\frac{d}{2}{F_p}_{\rho\sigma}{F_p}^{\rho\sigma}}\, \\
=&	\, F_p^{\mu\nu}\partial_{\nu}\leri{4b (\mathbf{E}_p^2-\mathbf{B}_p^2)
	+2d\, \mathbf{E}_p\cdot\mathbf{B}_p}
	+\tilde{F}_p^{\mu\nu}\partial_{\nu}\leri{c (\mathbf{E}_p^2-\mathbf{B}_p^2)
	+2d\, \mathbf{E}_p\cdot\mathbf{B}_p}\, ,
\end{split}
\end{equation}
where in the second line we rewrote $F_p^{\mu\nu}$ in terms of its electric and magnetic fields $\mathbf{E}_p$ and $\mathbf{B}_p$. 
As a result, the induced field, $F^{\mu\nu}$, is generated proportionally to the cubic power of the background fields. 
A similar effect occurs if the photon self-interactions are mediated by an off-shell scalar $\phi$ with the Lagrangian in Eq.~\eqref{eq:Lphi}. 
In this case, the effective current reads
\begin{align}
	\label{eq:Maxwell}
	J^\mu
=&	-gF_p^{\mu\nu}\partial_{\nu}\phi-\tilde{g}\tilde{F_p}^{\mu\nu}\partial_{\nu}\phi\,,
\end{align} 
where, at leading order in $g$ and $\tilde{g}$, $\phi$ is the solution of the Klein--Gordon equation $(\partial^2+m_\phi^2)\phi=J_\phi\equiv-\frac14 g{F_p}_{\mu\nu}{F_p}^{\mu\nu}-\frac14 \tilde{g}{F_p}_{\mu\nu}\tilde{F_p}^{\mu\nu}$, \ie~$\phi(t,\mathbf{x})=\int d^3 x dt \,G_R(\mathbf{x},\mathbf{x}',t,t') J_\phi(t,\mathbf{x}')$, where $G_R$ is the retarded Green's function. 
The calculation of $G_R$ and $\phi$ is simplified if $m_\phi$ is much larger or much smaller than the typical frequency of the background field $\omega$.
In these cases
\begin{align}
	\begin{cases}
	\phi_\infty
=	\frac{1}{m_\phi^2}\left[-\frac12 g\leri{\mathbf{B}_p^2-\mathbf{E}_p^2}+\tilde{g}\mathbf{B}_p\cdot\mathbf{E}_p\right] & m_\phi\gg\omega\\
	\phi_0
=	\frac{1}{4\pi }\int\frac{d^3x'}{|\mathbf{x}-\mathbf{x}'|}\left[-\frac12 g\leri{\mathbf{B}_p^2(t_R,\mathbf{x}')-\mathbf{E}_p^2(t_R,\mathbf{x}')}
	+\tilde{g}\mathbf{B}_p(t_R,\mathbf{x}')\cdot\mathbf{E}_p(t_R,\mathbf{x}')\right] & m_\phi \ll \omega
	\end{cases} \, ,
\end{align}
where $t_R\equiv t-|\mathbf{x}-\mathbf{x}'|$ is the retarded time. 
Note that the effective current in Eq.~\eqref{eq:Maxwell} is also cubic in the background fields and reduces to Eq.~\eqref{eq:JEFT1} in the limit $m_\phi\gg\omega$, with the Wilson coefficients given in Eq.~\eqref{eq:bcdphi}.

\begin{figure}[t]
	\centering
	\includegraphics[scale=0.25]{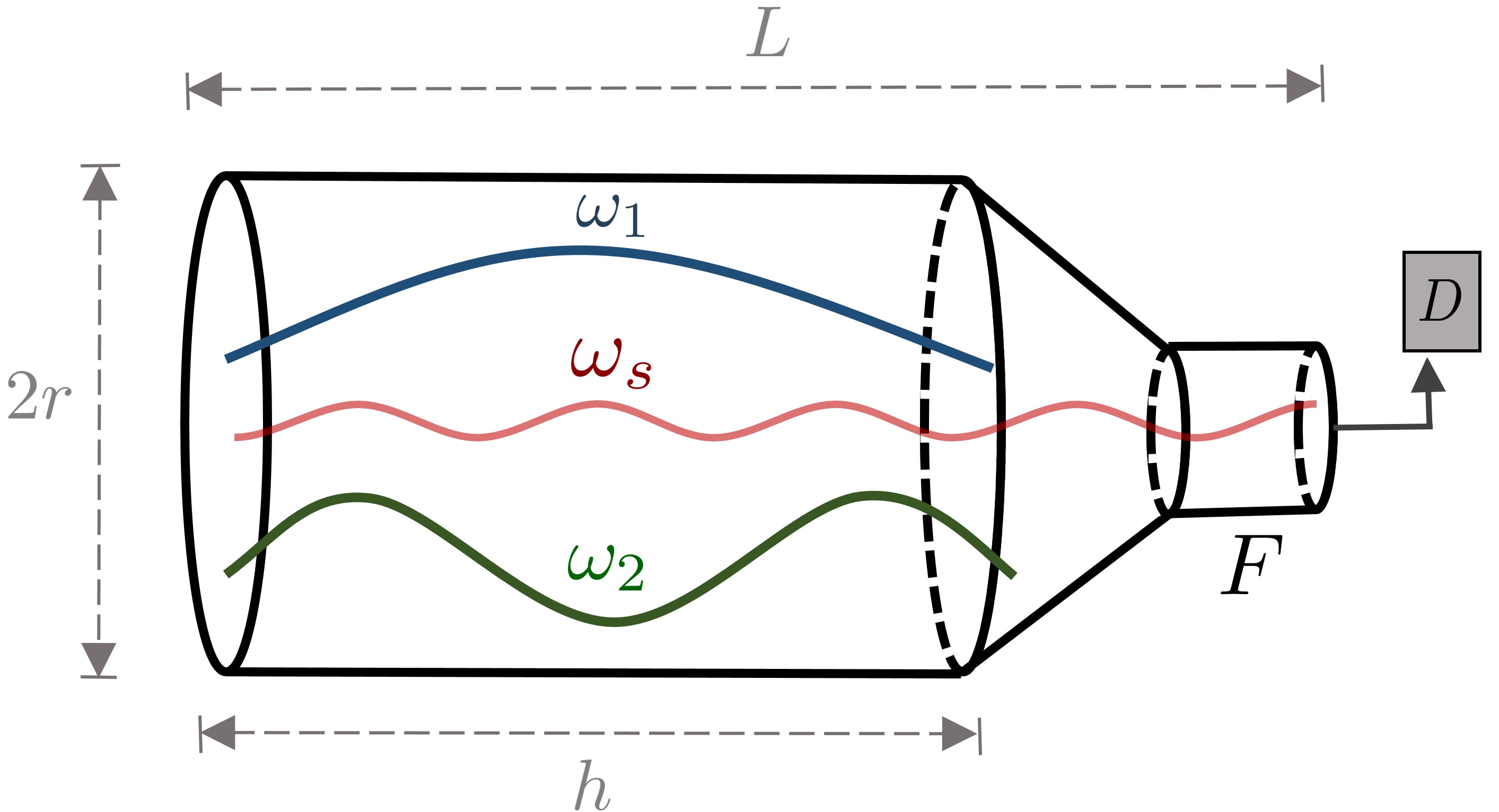}
	\caption{{\small Schematic picture of a cylindrical \ac{SRF} cavity. Two cavity modes with frequency $\omega_1$ and $\omega_2$ are pumped into the cavity and source an additional mode at the frequency $\omega_s>\omega_1,\omega_2$ as a result of the photon self-interactions of Eq.~\eqref{eq:LEFT}. The cavity geometry is chosen such that $\omega_s$ is a resonance mode and therefore amplified. In a small filtering region (F) the pump modes are exponentially suppressed and only the signal mode $\omega_s$ is detected (D). If the modes satisfy Eq.~\eqref{eq:conditions}, the signal mode, $\omega_s$, will be automatically sourced only by the \ac{CPV} part of the photon interactions (see main text for more details).}}
	\label{fig:SRF}
\end{figure}

As suggested in~\cite{Eriksson:2004cz}, an \ac{SRF} cavity is a natural setup where the field $F^{\mu\nu}$ can be generated and amplified. 
We consider an \ac{SRF} cavity, see Fig.~\ref{fig:SRF}, that is pumped simultaneously with two cavity modes, with corresponding electric fields $\mathbf{E}_1$, $\mathbf{E}_2$ and magnetic fields $\mathbf{B}_1$, $\mathbf{B}_2$, at frequencies $\omega_1$ and $\omega_2$ respectively, with $F^{\mu\nu}_p=F^{\mu\nu}_1+F^{\mu\nu}_2$ (we discuss the possibility of pumping the cavity with a single mode at the end of this section).
Since the modes of $F^{\mu\nu}$ that match resonances of the cavity are amplified by the cavity geometry, the electric field produced in the cavity will mostly be sourced by the projection of $\mathbf{J}$ onto these resonance modes. 
The resonant field $\mathbf{E}_{f}$ generated by exciting a cavity eigenmode $\hat{\mathbf{E}}_s$, with a corresponding frequency $\omega_s$, can be written as~\cite{Bogorad:2019pbu}
\begin{align}
	\label{eq:Ef}
	\mathbf{E}_{f}(\mathbf{x})
=	\frac{Q_s}{\omega_sV}\hat{\mathbf{E}}_s(\mathbf{x})\int d^3x'\,\hat{\mathbf{E}}_s(\mathbf{x}')\cdot \mathbf{J}(\mathbf{x}')\,,
\end{align}
where $V$ is the volume of the cavity, $Q_s$ is the quality factor for the frequency $\omega_s$ and $\hat{\mathbf{E}}_s$ is dimensionless and normalized as $\int d^3x |\hat{\mathbf{E}}_s|^2=V$.

Note that in order to excite the cavity resonance, one of the cavity resonance frequencies must match one of the Fourier components of $F^{\mu\nu}$. 
Given the cubic dependence of $\mathbf{J}$ on the pump fields, and assuming no other background sources, eq.~\eqref{eq:sourceEq} dictates that $F^{\mu\nu}$ can only have frequencies $\pm n\omega_1\pm m\omega_2$, with $m,n\geq0$ and $m+n=3$. 
The cavity geometry must be therefore chosen such that there exists a resonance frequency $\omega_s$ matching one of these combinations (within the cavity's bandwidth), while also verifying that the spatial overlap of the corresponding resonance mode $\hat{\mathbf{E}}_s$ and the effective current $\mathbf{J}$  (\ie~Eq.~\eqref{eq:Ef}) is non-zero (and ideally maximal).

To measure the power in the excited signal mode, one could use a smaller filtering cavity (of which $\omega_s$ is still a resonance mode), as is schematically shown in Fig.~\ref{fig:SRF}. Assuming $\omega_1,\omega_2<\omega_s$, this suppresses the pump fields and isolates $E_f$ only~\cite{Eriksson:2004cz}. 
The expected number of signal photons is given by
\begin{align}
	N_s
=	\frac{1}{2\omega_s}\int d^3x \left|\mathbf{E}_f(x)\right|^2 \, .
\end{align}

We will now show that, due to the intrinsically different structure of the \ac{CPC} and \ac{CPV} couplings, it is possible to select the pump fields $\mathbf{E}_1$ and $\mathbf{E}_2$, and the signal mode $\hat{\mathbf{E}}_s$, to single out \ac{CPV} (or \ac{CPC}) phenomena in the generated field of Eq.~\eqref{eq:Ef}. 

First, we observe that the effective current can be written as $J^\mu=F^{\mu\nu}\partial_\nu f+\tilde{F}^{\mu\nu}\partial_\nu \tilde{f}$, where $f$ and $\tilde{f}$ are quadratic functions of the pump fields and can be read off from Eqs.~\eqref{eq:JEFT1} and~\eqref{eq:Maxwell}. 
In particular, the vector current is $\mathbf{J}=(\mathbf{E}\partial_0+\mathbf{B}\cross\grad)f+(\mathbf{B}\partial_0-\mathbf{E}\cross\grad)\tilde{f}$. Let us consider a setup in which the signal mode $\hat{\mathbf{E}}_s$ is parallel to $\mathbf{E}_2$, the pump fields are orthogonally polarized, namely $\mathbf{E}_1\cdot\mathbf{E}_2=\mathbf{B}_1\cdot\mathbf{B}_2=\mathbf{E}_1\cdot\mathbf{B}_1=\mathbf{E}_2\cdot\mathbf{B}_2=0$, and are chosen such that either $\mathbf{B}_1\parallel \mathbf{E}_2$ or $(\grad f)_{\mathbf{B}_1\cross \mathbf{E}_2}=0$ (with the notation $(\mathbf{A})_{\mathbf{B}}$ we refer to the component of $\mathbf{A}$ along $\mathbf{B}$). 
With this choice, the scalar product entering in Eq.~\eqref{eq:Ef} simplifies considerably and reads
\begin{align}
	\label{eq:JEs}
	\hat{\mathbf{E}}_s\cdot\mathbf{J}
=	\hat{\mathbf{E}}_s\cdot\left[(\mathbf{E}_2\partial_0+\mathbf{B}_2\cross\grad)f
	+(\mathbf{B}_1\partial_0-\mathbf{E}_1\cross\grad)\tilde{f}\right]\,
\end{align}
The expressions for $f$ and $\tilde{f}$ are also simplified for the above choice of pump fields. From Eq.~\eqref{eq:JEFT1}, schematically $f\sim b F_p^2+dF_p\tilde{F}_p$ and $\tilde{f}\sim cF_p\tilde{F}_p+d F_p^2$, and the choice of orthogonal pump modes implies $F_p^2 \sim\mathbf{E}_1^2+\mathbf{E}_2^2-\mathbf{B}_1^2-\mathbf{B}_2^2$ and $F_p\tilde{F}_p \sim \mathbf{E}_1\cdot\mathbf{B}_2+\mathbf{B}_1\cdot\mathbf{E}_2$. Plugging these expressions into Eq.~\eqref{eq:JEs}, we see that the \ac{CPV} terms (proportional to $d$) contain an odd number of powers of the field `1' and an even number of `2', and vice-versa for the \ac{CPC} terms (proportional to $b$ or $c$). 
This happens because, crucially, only the fields $\mathbf{E}_1$ and $\mathbf{B}_1$ ($\mathbf{E}_2$ and $\mathbf{B}_2$) enter in the term operating on $\tilde{f}$ ($f$) in Eq.~\eqref{eq:JEs}, thanks to the properties of the pump modes. The same holds for the current in Eq.~\eqref{eq:Maxwell}, since only terms linear in $F^2$ and $F\tilde{F}$ can appear in $f$ and in $\tilde{f}$.

As a result, for modes that satisfy the conditions above, the \ac{CPV} part of the signal field $\mathbf{E}_f$ will only have frequency components
\begin{align}
	\label{eq:CPOfreq}
	\omega^{\rm CPV}_s 
=	\omega_1, \ 2\omega_2\pm \omega_1,  \ 3\omega_1	\, ,
\end{align}
\ie~combinations of odd multiplicities of $\omega_1$, and even multiplicities of $\omega_2$. 
The opposite holds for the \ac{CPC} terms, which only provide the frequency components 
\begin{align}
	\label{eq:CPEfreq}
	\omega^{\rm CPC}_s
=	\omega_2, \ 2\omega_1\pm \omega_2,  \ 3\omega_2 \,. 
\end{align}
Therefore, one can distinguish between \ac{CPV} and \ac{CPC} photon self-interactions according to the frequency of the component of $\mathbf{E}_f$.

As we are interested in observing and constraining \ac{CPV} photon self-interactions, we would like to amplify only the CP-odd contribution to field in the cavity (proportional to $d$ or $g\tilde{g}$). 
This can be done for the choice of pump fields proposed above, by setting the cavity geometry such that there exists a cavity eigenmode $\hat{\mathbf{E}}_s$, parallel to $\mathbf{E}_2$, with a frequency matching one of the possible CP-odd frequencies in Eq.~\eqref{eq:CPOfreq}. 
In particular, since for generic cavity geometries eigen-frequencies are not linear combinations of other eigen-frequencies, it should also be possible to make sure that the CP-even frequencies in Eq.~\eqref{eq:CPEfreq} do not match any of the resonant modes of the cavity (therefore preventing contaminations of the signal from the \ac{CPC} part). 
Summarizing, the conditions:

\begin{subequations}
	\label{eq:conditions}
	\centering
	\begin{tabularx}{\textwidth}{XX}
	\refstepcounter{mysub}\textup{(\themysub)}\ \ $\hat{\mathbf{E}}_s\parallel\mathbf{E}_2\,,$\\ 
	\refstepcounter{mysub}\textup{(\themysub)}\ \  
	$\mathbf{E}_1\cdot\mathbf{E}_2=\mathbf{B}_1\cdot\mathbf{B}_2=0$\,,
	&\hfill \textup{(\theequation)} \\
	\refstepcounter{mysub}\textup{(\themysub)}\ \ 
	$ \mathbf{B}_1\parallel\mathbf{E}_2$
	\label{eq:parallel}
	\ \   or  \ \   
	\refstepcounter{mysub}\textup{(\themysub)}
	\ \ $(\grad f)_{\mathbf{B}_1\cross \mathbf{E}_2}=0$
	\label{eq:grad}\,,\vfill
	\end{tabularx} 
\end{subequations}
together with the choice of $\omega_s$ among $\omega^{\rm CPV}_s$, are sufficient for isolating the \ac{CPV} part of the photon self-interactions from the \ac{CPC} one, as the signal will be affected only by the \ac{CPV} part.

Following the proposal in~\cite{Bogorad:2019pbu}, we now estimate the possible reach of the measurement of $d$, and of the \ac{CPV} combination $\sqrt{g\tilde{g}}$ of an off-shell scalar. 
We choose cavity modes satisfying Eq.~\eqref{eq:conditions} and normalize them as $\int d^3x |\mathbf{E}_1|^2=\int d^3x |\mathbf{E}_2|^2=E_0^2V$. 
We parametrize $\mathbf{J}=\kappa E_0^3\hat{\mathbf{J}}$, where $E_0$ is the typical magnitude of the electric field of the pump modes, $\hat{\mathbf{J}}$ is dimensionless and $\kappa$ has dimension $-3$. 
For the scalar mediator, we define $\kappa=\omega_s g\tilde{g}/m_\phi^2$ for $m_\phi\gg\omega_s$ and $\kappa=g\tilde{g}/\omega_s$ for $m_\phi\ll\omega_s$. 
When working in the \ac{EFT} limit, $\kappa=2\omega_s d$. 
The number of signal photons produced is then
\begin{align}
	\label{eq:NS1}
	N_s
=	\frac{Q^2_sE_0^6V}{2\omega^3_s}\kappa^2K^2\, , 
	\qquad\text{with} \qquad 
	K\equiv\frac1V\int d^3x \, 	\hat{\mathbf{E}}_s \cdot \hat{\mathbf{J}} \, ,
\end{align}
where $K$ is an $\mathcal{O}(1)$ form factor of the cavity that depends on the pump and signal modes, which will be computed in the following for explicit examples. We will use the notation $K_\infty$ when describing the \ac{EFT} limit, and $K_0$ in the $m_\phi\ll \omega_s$ case.

We obtain the expected bounds on the appropriate \ac{CPV} parameters for the experimental phases proposed in~\cite{Bogorad:2019pbu}, assuming thermal photons are the dominant noise source (in our analysis we neglect possible backgrounds from impurities of the cavities material, which can induce a noise in the signal mode, see~\cite{Gao:2020anb}). 
By following the procedure of Ref.~\cite{Bogorad:2019pbu}, the projected bound on \ac{CPV} combination of couplings of a scalar field to photons, $\sqrt{g\tilde{g}}$, is obtained by comparing the number of thermal photons and signal photons from Eq.~\eqref{eq:NS1} and is given by
\begin{align}
	\label{eq:ggtlim}
	\sqrt{g\tilde{g}}^{\rm lim.} 
	\sim 
	\left(\frac{4T L}{Q V E_0^6} \sqrt{\frac{B_\omega}{t}} \text{SNR} \right)^{\frac14} \times	
	\begin{cases}
		K_0^{-1/2} \omega_s ,  & m_\phi \ll \omega_s\,,\\
		K_\infty^{-1/2} m_\phi,  & m_\phi \gg \omega_s \,,
	\end{cases}
\end{align}
where $T$ is the cavity's temperature (we assume $T \gg \omega_s$), $t$ is the total measurement time, $B_\omega$ is the signal bandwidth, $L$ is the total length of the cavity (including the filtering region), and $\text{SNR}$ is the \acl{SNR}. 
We set $\text{SNR}=2$ to obtain the bound at a $95\%$ confidence level. 
The projected bounds on the coefficient $d$ of the generic \ac{EFT} in Eq.~\eqref{eq:LEFT} is
\begin{align}
	\label{eq:dlimSRF}
	d^\text{lim}
=	\frac{1}{K_\infty}\sqrt{\frac{TL}{Q_sV E_0^6}\sqrt{\frac{B_\omega}{t}}\text{SNR}}\,.
\end{align}
\begin{table}[t]
\begin{center}
\begin{tabular}{|c|c|c|c|c|c|c|}
\hline
phase & $r\,$[m] & $h\,$[m] & $\omega_s\,$[GHz] & $Q_s$ & $B_\omega\,$[Hz] & $t\,$[days] \\
\hline\hline
1\,(a) &  $0.5$ & $0.166$ & $10.84$ & $2.6 \times 10^8$ & 2  & 1 \\
\hline
1\,(b) &  $0.5$ & $0.166$ & $10.84$ & $2.6\times 10^8$ & $1/t$  & 1\\
\hline
2 &  $0.5$ & $0.166$ & $10.84$ & $ 10^{12}$ & $1/t$  & 20\\
\hline
3 &  $2$ & $0.664$ & $2.71$ & $ 10^{12}$ & $1/t$  & 365\\ 
\hline
\end{tabular}
\end{center}
\caption{The parameters the different phase of the \ac{SRF} cavities. 
The modes pumped in the cavity are assumed to be $\text{TE}_{021}/\text{TM}_{050}/\text{TM}_{060}$. For all phases we assume $T=1.5\,$K and $E_0=45$\,MV/m.}
\label{tab:phases}
\end{table}
\begin{figure}[t]
	\centering
	\includegraphics[scale=0.5]{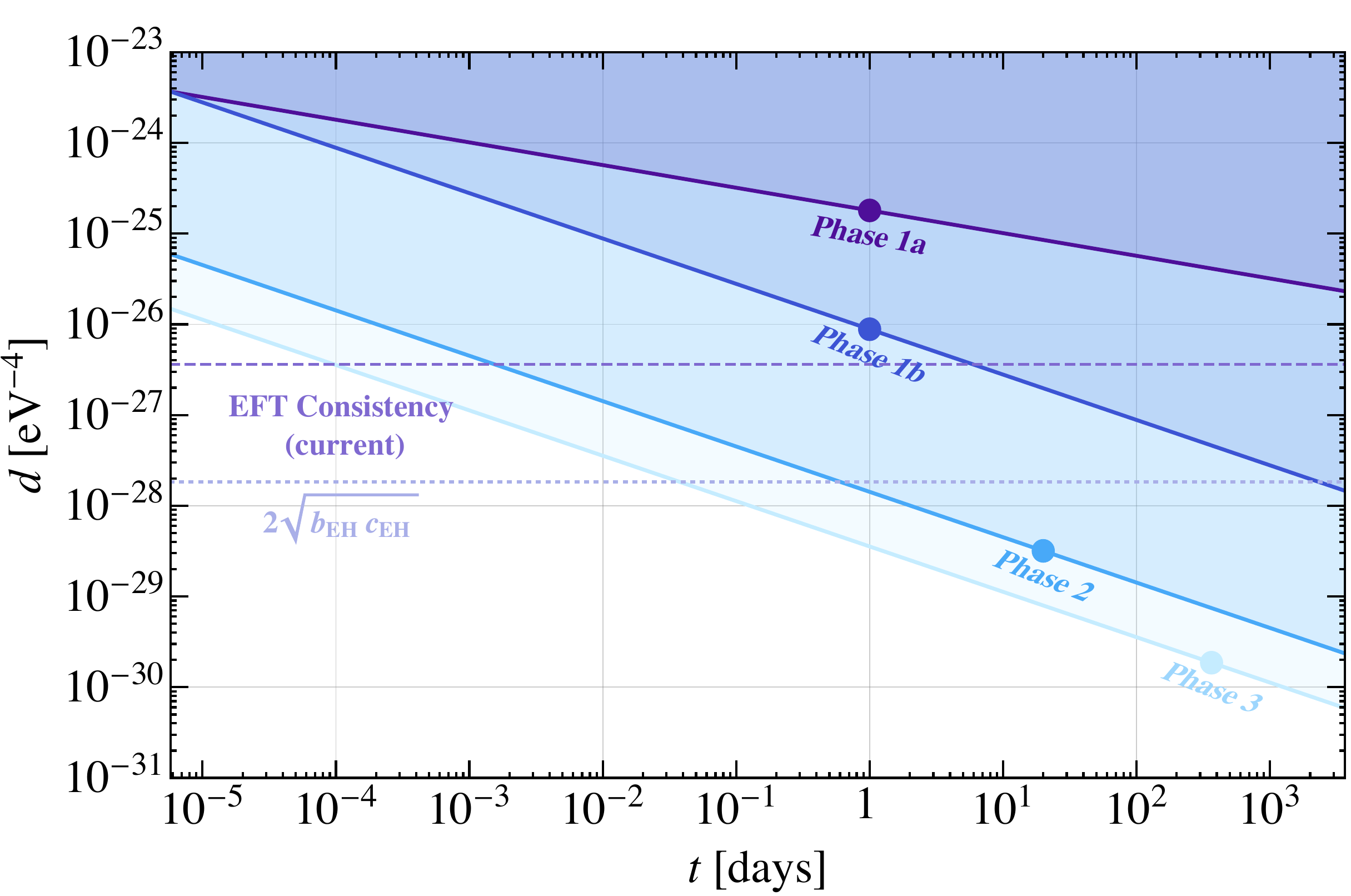}
	\small{\caption{Prospective bound on the coefficient $d$ of the effective operator $FFF\tilde{F}$ of Eq.~\eqref{eq:LEFT} 
	as a function of the total measurement time in the \ac{SRF} cavity \,(solid lines). 
	The different lines correspond to the typical cavity parameters of the phases in proposed in~\cite{Bogorad:2019pbu}, summarized in Table~\ref{tab:phases}. The disk shows the total measurement time proposed in~\cite{Bogorad:2019pbu}.
	The upper dashed line is the consistency bound $d\leq2\sqrt{bc}$ for the photon \ac{EFT} from Eq.~\eqref{eq:EFTcon}, with the current direct experimental limits on $b$ and $c$ 
	(see Eqs.~\eqref{eq:bLambShift} and~\eqref{eq:cComb}). The lower dashed line corresponds to $d\leq2\sqrt{b_\text{EH}c_\text{EH}}$, which would be the consistency bound on the \ac{CPV} coefficient assuming the \ac{EH} effect would be measured, and thus represents the equivalent sensitivity to the \ac{EH} scale.
		\label{fig:SRFboundsd} 
	}}
\end{figure}
\begin{figure}[t]
	\centering
\includegraphics[scale=0.5]{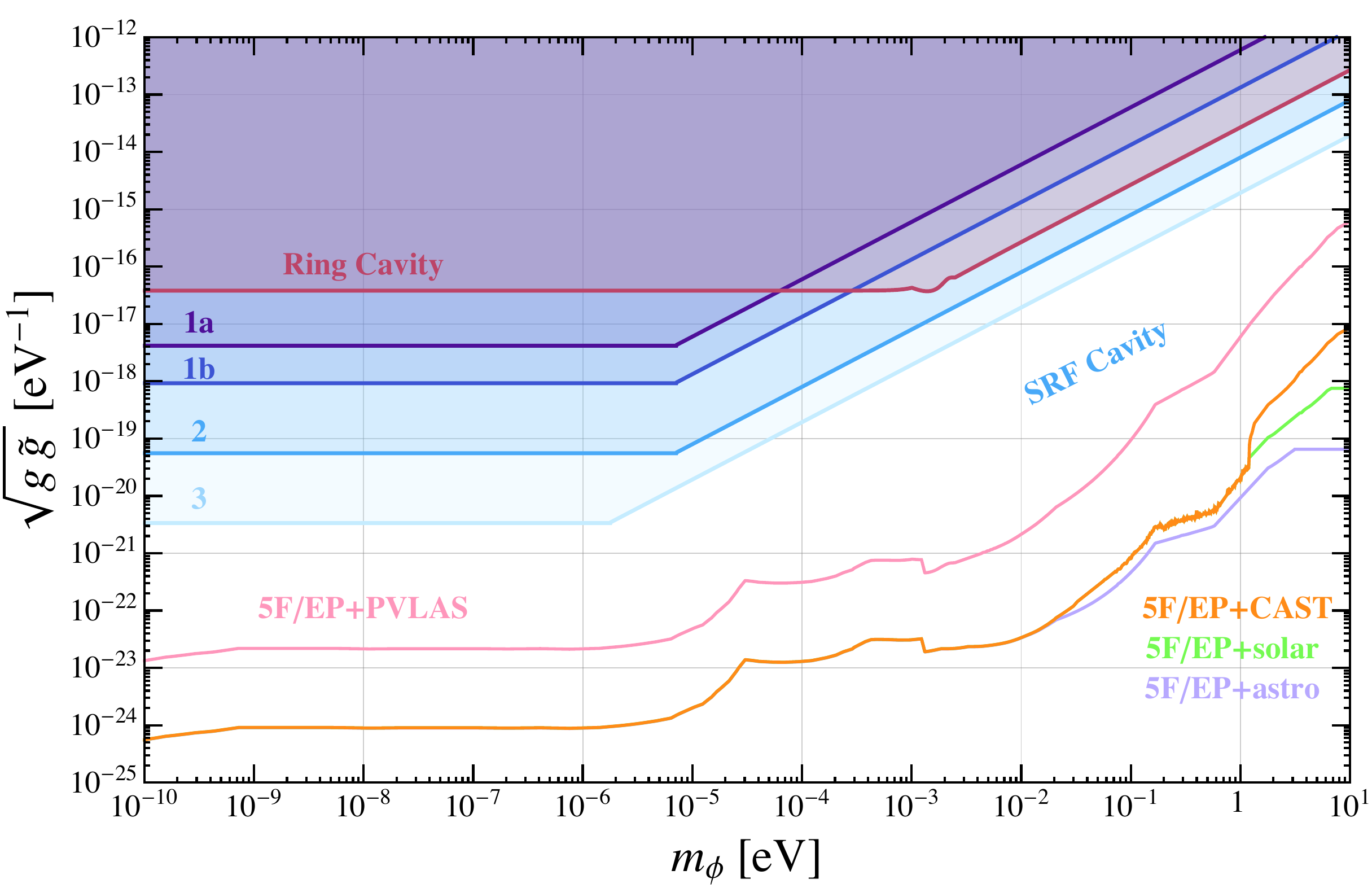}
	\caption{Projected sensitivity to the CP-violating combination $\sqrt{g\tilde{g}}$ of the couplings of a scalar to photons, with the Lagrangian in Eq.~\eqref{eq:Lphi}, as a function of the mass $m_\phi$. 
		We show in blue bounds expected for the proposal presented in Sec.~\ref{sec:SRF} (SRF cavity), with the experimental parameters and the integrated measurement times given in Table~\ref{tab:phases}.
		In red we show the bounds for the proposal presented in Sec.~\ref{sec:RingCavity} (ring cavity), for the same experimental parameters as PVLAS. 
		For reference, we also plot the current best bound on $\sqrt{g\tilde{g}}$, obtained by combining the constraints on $g$ from fifth force~\cite{Adelberger:2003zx,Lee:2020zjt} 
		and \ac{EP} tests~\cite{PhysRevLett.120.141101,PhysRevLett.100.041101,Wagner2012,PhysRevD.61.022001} (summarized in~\cite{Hees:2018fpg}) (5F/EP), 
		and those on $\tilde{g}$ from PVLAS/CAST~\cite{Ejlli:2020yhk,Anastassopoulos:2017ftl} and astrophysics (see \eg~\cite{Irastorza:2018dyq}).}
	\label{fig:SRFboundsggt}
\end{figure}

As a proof of principle, we consider a cavity with a right cylindrical geometry with radius $r=0.5$~m (see Fig.~\ref{fig:SRF}). 
The conditions in Eq.~\eqref{eq:conditions} can be satisfied, for example, for the orthogonal pump fields of the form $F_1=\text{TE}_{0q_1p_1}$ and $F_2=\text{TM}_{0p_20}$, with the signal mode $F_s=\text{TM}_{0p_s0}$ (see \eg~\cite{Hill:2014} for notation and analytic expressions). 
With this choice, $\hat{\mathbf{E}}_s$ is parallel to $\mathbf{E}_2$ and points to the $z$ direction. 
For these modes, $\mathbf{B}_1\times\hat{\mathbf{E}}_s\propto\hat{\mathbf{\varphi}}$, where $\varphi$ is the azimuthal angle. 
Since the system (fields and boundary conditions) is azimuthally symmetric, condition~\eqref{eq:grad} of Eq.~\eqref{eq:conditions} is also satisfied. 
Another class of modes satisfying Eq.~\eqref{eq:conditions} is $\text{TM}_{0p_10}/\text{TE}_{0p_2q_2}/\text{TE}_{0p_sq_s}$. 
As in the previous mode choice, the pump fields are orthogonal and $\hat{\mathbf{E}}_s\parallel \mathbf{E}_2$. 
However, for this choice, $\mathbf{B}_1\parallel \mathbf{E}_2$, satisfying condition~\eqref{eq:parallel}.
In Tables~\ref{modes-TM} and~\ref{modes-TE} of Appendix~\ref{sec:cavitymodes} we list mode combinations of the form $\text{TE}_{0q_1p_1}/\text{TM}_{0p_20}/\text{TM}_{0p_s0}$ and $\text{TM}_{0p_10}/\text{TE}_{0p_2q_2}/\text{TE}_{0p_sq_s}$ that yield $\mathcal{O}\leri{0.1-1}$ values for $K_\infty$ and $K_0$, for $\omega_s=2\omega_2-\omega_1$.

In the following we will use Eqs.~\eqref{eq:ggtlim}--\eqref{eq:dlimSRF} to estimate 
the projected bounds on $d$ and $g\tilde{g}$ achievable using our proposed method.
We choose the $\text{TE}_{021}/\text{TM}_{050}/\text{TM}_{060}$ modes configuration, representing the first type of mode combinations mentioned above. 
As discussed before, by choosing the cavity geometry such that $\omega_s=2\omega_2-\omega_1$, \ie~$\omega_{\text{TM}_{060}}=2\omega_{\text{TM}_{050}}-\omega_{\text{TE}_{021}}$, only the \ac{CPV} part of the Lagrangian will contribute to $\mathbf{E}_f$. 
This is achieved if the cavity length is $h=0.332r=0.166\,$m. 
For this mode choice, we find $K_\infty= 0.25$ and $K_0= 0.26$. 
We set the total length of the cavity with the filtering region to be $L=2h$, and assume the radius of the filtering cavity will be chosen such that $\mathbf{E}_f$ corresponds to the lowest resonance mode of the filtering cavity. 
We give our projections for four cases, following the four phases of Ref.~\cite{Bogorad:2019pbu}, where the different operating parameters are given in Table~\ref{tab:phases}.
For all cases we assume that the cavity temperature is $T=1.5\,$K and that $E_0=45$\,MV/m.

The corresponding expected bound on the \ac{EFT} coefficient $d$ is plotted in Fig.~\ref{fig:SRFboundsd}. We observe that such a cavity can easily probe values of $d$ that are within the \ac{EFT} consistency region given by Eq.~\eqref{eq:EFTcon}, providing the first direct limit on this coefficient to date. In particular, the reach could even be a few orders of magnitude below the value $2\sqrt{b_\text{EH}c_\text{EH}}\simeq 0.24\alpha_{\rm EM}^2/m_e^4$, which corresponds to the \ac{EFT} consistency scale if $b$ and $c$ are measured to be at the values predicted by the \ac{EH} effect, and thus represents the equivalent effective sensitivity associated with the \ac{EH} scale. The projected sensitivity for the \ac{CPV} combination of scalar couplings $\sqrt{g\tilde{g}}$, presented in Fig.~\ref{fig:SRFboundsggt}, is still however a few order of magnitudes above the current best bounds on these couplings, but would provide a strong complementary probe. Additionally, when applied to the fermion dipole operators in Eq.~\eqref{eq:Lpsi}, our best bound on $d$ would constrain $D\,,\tilde{D}\lesssim 5\times 10^{-8}\,\eV^{-1}$, even in the mass range $\text{few  }100\,\keV \lesssim m_\psi\lesssim 10\,\MeV$  which is presently not directly constrained by astrophysical observations.

We note that it is possible to disentangle the contribution of the \ac{CPC} and \ac{CPV} coefficients using just a single pump mode (that self-interacts with itself) with orthogonal electric and magnetic fields, \ie~$\mathbf{E}_p\cdot\mathbf{B}_p=0$.
In this case the scalar product in Eq.~\eqref{eq:Ef} is
\begin{align}
\mathbf{\hat{E}}_s\cdot\mathbf{J}=\mathbf{\hat{E}}_s\cdot[(\mathbf{E}_p\partial_0+\mathbf{B}_p\cross\grad)f+(\mathbf{B}_p\partial_0-\mathbf{E}_p\cross\grad)\tilde{f}],
\end{align} 
where $f\sim b (\mathbf{E}_p^2-\mathbf{B}_p^2)$ and $\tilde{f}\sim d (\mathbf{E}_p^2-\mathbf{B}_p^2)$ given that $F_p\tilde{F}_p=4\mathbf{E}_p\cdot\mathbf{B}_p=0$. 
Therefore if $\mathbf{\hat{E}}_s$ is chosen to be parallel to $\mathbf{E}_p$ ($\mathbf{B}_p$) only the term containing $f$ ($\tilde{f}$) survives in $\mathbf{\hat{E}}_s\cdot\mathbf{J}$ and the signal will be affected only by $b$ ($d$). 
In both cases, the cavity dimensions should be chosen such that the signal resonance mode satisfies $\omega_s=3\omega_p$ in order for the signal be amplified and isolated. 
This method is however unable to constrain effective CP-even interactions of the form $(F\tilde{F})^2$ (as could be generated by an \ac{ALP}), since it requires $F_p\tilde{F}_p\neq0$.
 For concreteness, we have tested the \ac{CPV} combination $F_p=\text{TM}_{010}$ and $F_s=\text{TE}_{011}$, and found $K_\infty\sim0.09$. It is possible that higher form factors are attainable, although they may be harder to optimize, as suggested by the authors of~\cite{
Eriksson:2004cz}.

\section{CP-violation and Vacuum Birefringence}
\label{sec:RingCavity}

In this section, we discuss the effect of CP-odd photons self-interactions on the birefringence properties of the vacuum. In particular, we will show that a setup where vacuum birefringence takes place in a ring cavity is sensitive to both \ac{CPC} and \ac{CPV} phenomena separately.

\begin{figure}[t]
	\centering
	\includegraphics[width=0.45\textwidth]{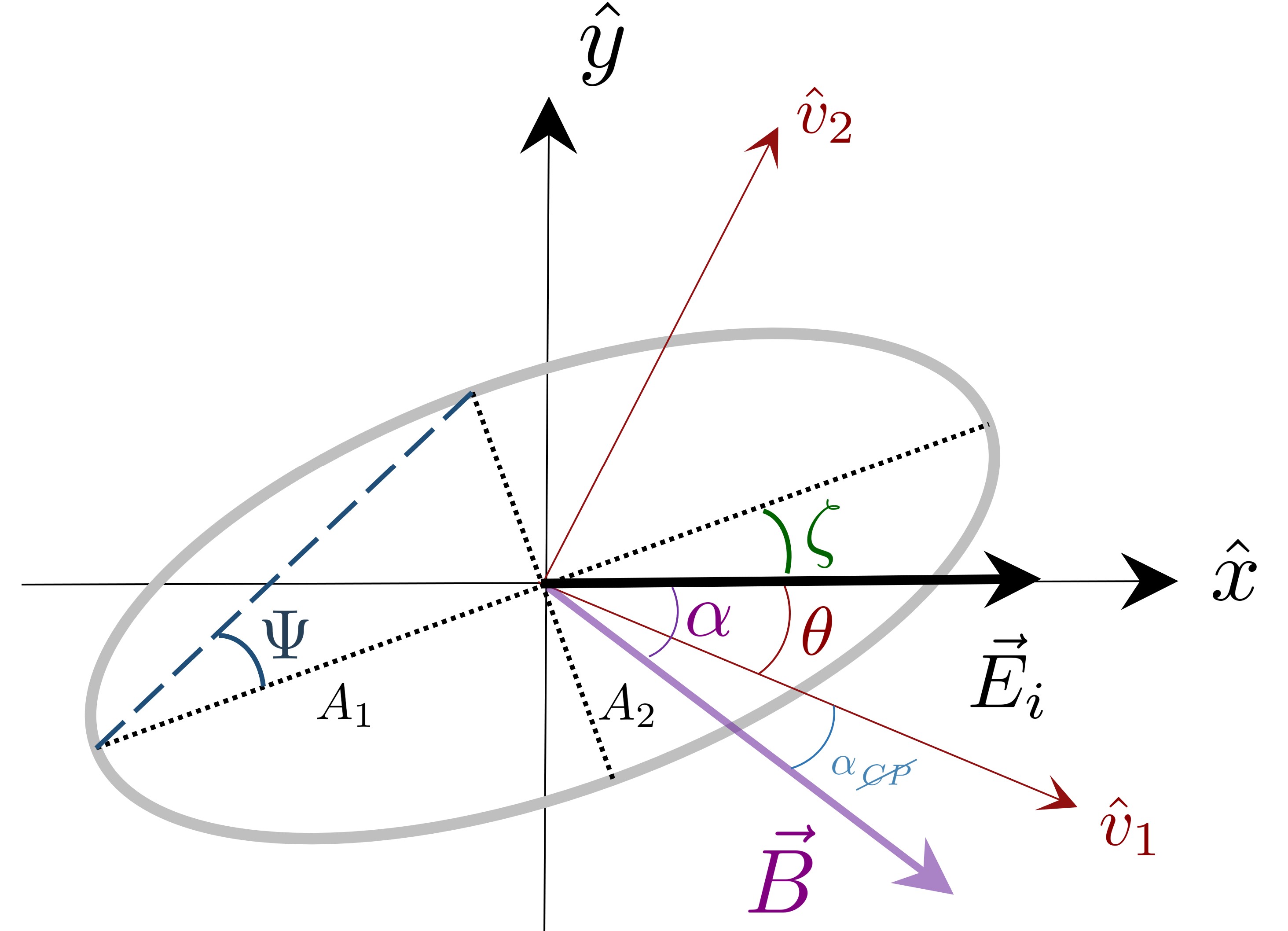}	
	\caption{{\small In the presence of photon self-interactions, an initially linearly polarized light (traveling along the $\hat{z}$ direction, perpendicular to the page) with polarization vector $\mathbf{E}_i$ acquires an elliptical polarization and a rotation of the polarization plane after propagating an effective distance $L/\lambda$ through a region permeated by an external magnetic field $\mathbf{B}$ (parallel to the $x-y$ plane). The ellipticity $\Psi$ and rotation $\zeta$ are defined as in the picture (see main text for more details).}}
	\label{fig:PVLSlike}
\end{figure}

Nonlinearities in Maxwell's equations are known to introduce a nontrivial response of the vacuum in the presence an external electromagnetic field~(see \eg~\cite{ADLER1971599,Liao:2007nu,PhysRevD.75.117301,Millo:2008ug,Battesti:2018bgc}). 
Consider a light beam with a frequency $\omega$, linearly polarized along the $\hat{x}$ axis  and propagating along the $\hat{z}$ direction through a region of length $L$ permeated by an external static magnetic field $\mathbf{B}$, which lies in the $x-y$ plane (see Fig.~\ref{fig:PVLSlike}). 
The external magnetic field induces two different refractive indices, $n_1$ and $n_2$, along the two orthogonal 
directions $\hat{v}_1$ and $\hat{v}_2$, shown in Fig.~\ref{fig:PVLSlike}. 
The magnitude of $n_1$ and $n_2$, and the orientation of the axes $\hat{v}_1$ and $\hat{v}_2$, are determined by $\mathbf{B}$ and the photon self-interactions.

As a result of the anisotropic refractive index, the components of the electric field $\mathbf{E}_i$ of the beam along $\hat{v}_1$ and $\hat{v}_2$ evolve separately, inducing a change in the polarization vector the probe
~\cite{Liao:2007nu,Millo:2008ug}. Below we consider two observables that are sensitive to an anisotropic refractive index, both of which, as we will see, can be split into the sum of CP-even and CP-odd components.

First, the propagation inside the birefringent region will cause the linearly polarized beam to become elliptically polarized, as the projections of the field onto $\hat{v}_1$ and $\hat{v}_2$ propagate at two different velocities ($1/n_1$ and $1/n_2$), and thus acquire a phase difference $\delta\varphi$ which varies along the propagation distance as $\delta\varphi=\leri{n_1-n_2}\omega z$. This can be easily seen by noticing that the evolution of the polarization vector is given by
\begin{align}\label{eq:Etx}
\mathbf{E}(t,\mathbf{x})=\exp\left(i\omega t\right)\left[\hat{v}_1 \exp(-i\omega n_1 \mathbf{\hat{k}}\cdot \mathbf{z})\,\hat{v}_1\cdot \mathbf{E}_i+\hat{v}_2 \exp(-i\omega n_2 \mathbf{\hat{k}}\cdot \mathbf{z})\,\hat{v}_2\cdot \mathbf{E}_i\right]\,,
\end{align}
where $\hat{\mathbf{k}}$ is the propagation direction of the beam, which we assume is either parallel or anti-parallel to $\hat{z}$.
The relative phase between the two components of the field is usually quantified in terms of the ellipticity $\Psi$, which is defined by the ratio $A_1/A_2$ of the axes of the polarization ellipse via $\tan\Psi \equiv A_2/A_1$ (see Fig.~\ref{fig:PVLSlike}). Equivalently, in terms of the electric field in Eq.~\eqref{eq:Etx}, we may define $\sin2\Psi=i\mathbf{\hat{k}}\cdot (\mathbf{E}\times \mathbf{E^*})/|\mathbf{E}|^2$~\cite{BornMax1980PoOE}. 
As can be seen from Eq.~\eqref{eq:Etx}, if the eigen-axis $\hat{v}_1$ forms an angle $\theta$ with respect to the initial direction of polarization $\hat{\mathbf{E}}_i$, then in the limit $\Delta n\equiv n_2-n_1\ll1$ the ellipticity acquired over a distance $L$ is given by (see also \eg~\cite{Ejlli:2020yhk})
\begin{align}\label{eq:Psi}
\Psi=\frac{\omega L}{2}\Delta n\sin2\theta\,.
\end{align}

Second, as we will see shortly, in the presence of photon interactions with a light scalar, the refractive index can acquire a anisotropic imaginary part, expressed in terms of the absorption coefficients $\kappa_1\equiv\text{Im}[n_1]$ and $\kappa_2\equiv\text{Im}[n_2]$. From Eq.~\eqref{eq:Etx}, this corresponds to an anisotropic attenuation of the field, which can be interpreted as the decay of one component of the photon field into the on-shell scalar. In particular, if $\kappa_1\neq\kappa_2$ the components of the polarization vector along $\hat{v}_1$ and $\hat{v}_2$ are depleted differently, and the polarization ellipse rotates by an angle $\zeta$, defined as the angle between the major axis of the ellipse and the initial polarization direction $\mathbf{E}_i$ (see Fig.~\ref{fig:PVLSlike}). 
In the limit $\Delta \kappa \equiv \kappa_2-\kappa_1\ll1$, the acquired rotation over a propagation length $L$ is given by ~(see also \eg \cite{Ejlli:2020yhk})
\begin{align}\label{eq:zeta}
\zeta =\frac{\omega L}{2}\Delta \kappa\sin2\theta\,.
\end{align}

In the following, we will specialize the expressions of the ellipticity and rotation (written before for generic refractive indices and $\hat{v}_1$, $\hat{v}_2$) to those induced by photon interactions in the background of a magnetic field, mediated either by a light scalar with the Lagrangian in Eq.~\eqref{eq:Lphi} or via the effective interactions in Eq.~\eqref{eq:LEFT}. 
We will then show that both the ellipticity and the rotation can be broken into a CP-even and a CP-odd component, denoted by 
\begin{align}
	\label{eq:scalar_birefringence}
	\Psi
= 	\Psi^e + \Psi^o \, ,\qquad\quad
	\zeta
=	\zeta^e + \zeta^o \, ,
\end{align}
where the superscript $e\,(o)$ is for the CP even\,(odd) component. 

If the photon self-interactions are mediated by the effective Lagrangian in Eq.~\eqref{eq:LEFT}, the refractive indices $n_1$ and $n_2$ are real and read (see Appendix~\ref{sec:ellipt} for the explicit derivation)
\begin{align}\label{eq:n12}
n_{1,2}&=1+\frac{B^2}{2}\leri{4b+c\mp\sqrt{\leri{c-4b}^2+4d^2}}\,.
\end{align}
Note that the difference in refractive indices is proportional to the square of the magnetic field $\mathbf{B}$ and to the strength of the photon self-interactions. 

As mentioned earlier, the directions $\hat{v}_1$ and $\hat{v}_2$ are related to $\mathbf{B}$ and to the photon-self interaction coefficients. 
In the absence of CP violation, \ie~for $d=0$, it is easy to show that $\hat{v}_1$ coincides with $\hat{\mathbf{B}}$.
In this case, the orthogonal direction $\hat{v}_2$ is given by $\hat{\mathbf{k}}_i\cross\hat{\mathbf{B}}$, where $\mathbf{k}_i$ is the momentum of the initial beam. Note that we define the positive direction of all angles in the polarization plane as $\hat{\mathbf{B}}\times\hat{\mathbf{k}}_i\times\hat{\mathbf{B}}=\hat{\mathbf{k}}_i$.
Therefore, $\theta$ coincides with the angle $\alpha$ between $\mathbf{B}$ and $\mathbf{E}_i$.  
Instead, for $d\neq0$, $\{\hat{v}_1,\hat{v}_2\}$ are rotated with respect to $\{\hat{\mathbf{B}},\hat{\mathbf{k}}_i\times\hat{\mathbf{B}}\}$ by the angle $\alpha_{\cancel{\rm CP}}$ given by (see Appendix~\ref{sec:ellipt})
\begin{align}
	\label{eq:alphacpv}
 	\tan{(2\alpha_{\cancel{\rm CP}})}&=-(\hat{\mathbf{k}}_i \cdot \hat{\mathbf{k}})\frac{2d}{4b-c}\,.
\end{align}
As a consequence, if CP is broken, $\theta$ and $\alpha $ are different and related by $\theta=\alpha-\alpha_{\cancel{\rm CP}}$, as shown in Fig.~\ref{fig:PVLSlike}. 
In particular, according to our convention, the positive direction of $\alpha_{\cancel{\rm CP}}$ is constant and set by $\mathbf{k}_i$. 
As was pointed out in~\cite{Millo:2008ug}, flipping the propagation direction while keeping the polarization vector constant -- as is the case upon reflection off of a mirror in a zero incidence angle -- is equivalent to flipping the sign of the \ac{CPV} spurion $d$ and therefore the sign of $\alpha_{\cancel{\rm CP}}$ from Eq.~\eqref{eq:alphacpv}, as can be seen by the $(\hat{\mathbf{k}}_i \cdot \hat{\mathbf{k}})$ term (see also Eq.~\eqref{eq:nMatrixEFT} of Appendix~\ref{sec:ellipt}, which is invariant under $\mathbf{k}\to-\mathbf{k}$ and $d\to-d$).
This is a direct consequence of parity violation, which is equivalent to CP violation since charge conjugation is a symmetry of electrodynamics in vacuum.

Plugging Eq.~\eqref{eq:n12} and $\theta=\alpha-\alpha_{\cancel{\rm CP}}$ into the general expressions for the ellipticity and the rotation in Eq.~\eqref{eq:Psi} and Eq.~\eqref{eq:zeta} respectively, we find that the rotation vanishes (\ie~$\zeta_{\rm EFT}=0$, as was also shown in~\cite{PhysRevD.75.117301,Liao:2007nu}), and the ellipticity has the finite value
\begin{align}
	\label{eq:elliptEFT}
	\Psi_{\rm EFT}
=	\Psi_{\rm EFT}^e + \Psi_{\rm EFT}^o
=	\sin (2 \alpha -2 \alpha_\text{\cancel{CP}} )\frac{\omega L}{2}B^2\sqrt{(c-4 b)^2+4 d^2} \,.
\end{align}
As anticipated, the ellipticity in Eq.~\eqref{eq:elliptEFT} can be broken into the CP-even and CP-odd parts
\begin{align}
	\label{eq:elliptEFTCPE}
	\Psi_{\rm EFT}^e
=	(c-4 b)\frac{\omega L}{2}B^2 \sin\left( 2 \alpha\right) , \qquad 
	\Psi_{\rm EFT}^o
=	-2 (\hat{\mathbf{k}}_i \cdot \hat{\mathbf{k}}) d\frac{\omega L}{2}B^2 \cos\left( 2 \alpha\right) \, .
\end{align}

Similarly, if the photon interacts with a light scalar field with the couplings in Eq.~\eqref{eq:Lphi}, the acquired ellipticity $\Psi_\phi$ and rotation $\zeta_\phi$ are given by (in the small $\Psi_\phi$ and $\zeta_\phi$ limit, see~\cite{Raffelt:1987im,Redondo:2008tq,Liao:2007nu} for the explicit derivation)
\begin{align}
	\label{eq:elliptphi}
	\Psi_\phi
&=	B^2 \sin (2 \alpha -2 \alpha_{\cancel{\rm CP}} ) \frac{\omega  L}{4}
	\frac{(g^2+\tilde{g}^2)}{m_\phi^2}\left(1-\frac{\sin x}{x}\right)\,,\\	
\label{eq:rotphi}
	\zeta_\phi
&=	B^2 \sin (2 \alpha -2 \alpha_{\cancel{\rm CP}} ) \, \omega^2 
	\frac{(g^2+\tilde{g}^2)}{m_\phi^4}\sin^2\left(\frac{x}{2}\right) \,,
\end{align}
where $x\equiv m_\phi^2 L /(2\omega)$. In this case $\tan{\alpha_{\cancel{\rm CP}}}=\leri{\hat{\mathbf{k}}_i\cdot \hat{\mathbf{k}}} g/\tilde{g}$, and as before it changes sign under $\mathbf{k}\to-\mathbf{k}$. 
The ellipticity and the rotation in Eq.~\eqref{eq:elliptphi} and Eq.~\eqref{eq:rotphi} can again be broken into the CP-even and CP-odd parts
\begin{align}\label{eq:elliptphiCPE}
	\Psi_{\phi}^e
=& 	(\tilde{g}^2\!-\!g^2)\sin\left( 2\alpha\right)\left(1\!-\!\frac{\sin x}{x}\right)\frac{\omega B^2 L}{4m_\phi^2}	\, , 
&	\Psi_{\phi}^o
=&	-2 (\hat{\mathbf{k}}_i \cdot \hat{\mathbf{k}})g\tilde{g}\cos\left( 2\alpha\right)\left(1\!-\!\frac{\sin x}{x}\right)\frac{\omega B^2L}{4m_\phi^2}	\, , \\
	\zeta_\phi^e 
=&	(\tilde{g}^2\!-\!g^2)\frac{\omega^2 B^2}{m_\phi^4}\sin\left( 2\alpha \right) \sin^2\left( \frac{x}{2}\right)\, , 
&	\zeta_\phi^o 
=&	-2 (\hat{\mathbf{k}}_i \cdot \hat{\mathbf{k}}) g\tilde{g}\frac{\omega^2 B^2}{m_\phi^4}\cos\left( 2\alpha\right) \sin^2\left( \frac{x}{2}\right)\,.\label{eq:rotphiCPE}
\end{align}
Note that in the limit $m_\phi/\omega\to\infty$ the scalar expression in Eq.~\eqref{eq:elliptphi} consistently reproduces the expression for the \ac{EFT} in Eq.~\eqref{eq:elliptEFT} with the Wilson coefficients in Eq.~\eqref{eq:bcdphi}, while $\zeta_\phi$ as expected vanishes.

We note that the CP-even and CP-odd parts of the ellipticity and rotation in Eqs.~\eqref{eq:elliptEFTCPE},~\eqref{eq:elliptphiCPE} and~\eqref{eq:rotphiCPE} have different dependencies on the angle $\alpha$, and can therefore be studied separately. 
However, as already noticed also in Refs.~\cite{Liao:2007nu,Millo:2008ug,Redondo:2008tq,Fan:2017sxk}, while the former does not depend on the relative direction of propagation of the beam, the latter changes signs after switching the direction of propagation. 
Therefore, if the light beam is reflected within a region permeated by a magnetic field, the total change in ellipticity and rotation after a single round trip is (for perfect mirrors) respectively $2\Psi^e$ and $2\zeta^e$, and the CP-odd part cancels out. 
As a result, after $N$ trips, the total ellipticity increases by a factor of $(2N+1)$ only in its CP-even part, while remains unchanged in its CP-odd part. 
A setup involving multiple reflections can be therefore thought of as an optical path multiplier affecting only the CP-even part of $\Psi$ and $\zeta$.  
In the PVLAS experiment~\cite{DellaValle:2015xxa} the CP-even signal is enhanced in this way inside a linear \ac{FP} cavity. 
Since there is no amplification of the CP-odd component, the resulting sensitivity to CP-odd photon self-interactions is very weak.

\begin{figure}
	\centering
	\includegraphics[scale=0.22]{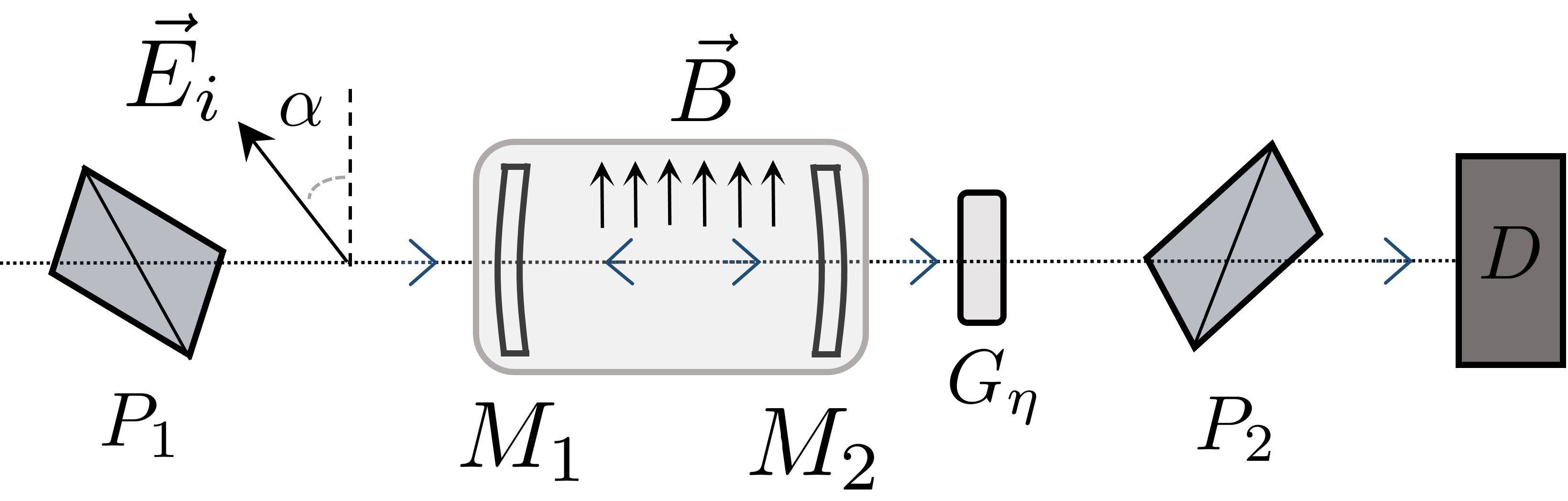}\qquad\qquad	
	\includegraphics[scale=0.23]{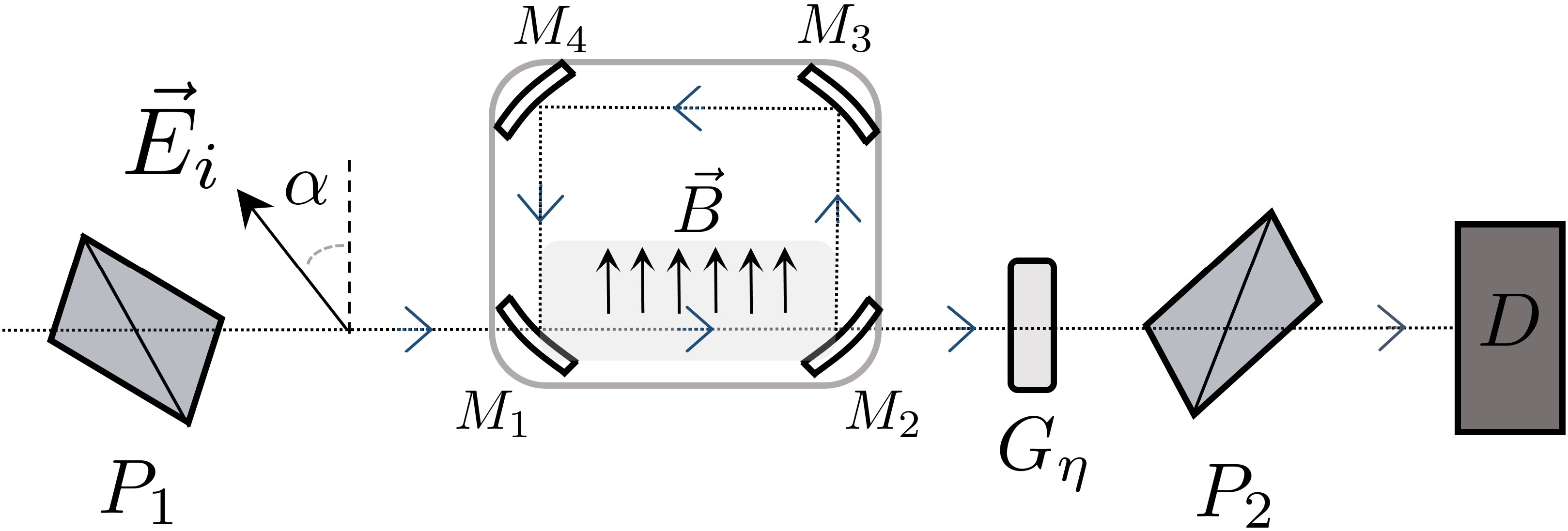}
	\vspace{2mm}
	\caption{{\small 
	\emph{Left}: Schematics of the PVLAS experiment. Linearly polarized light produced by the polarimeter $P_1$ enters the Fabry-Perot cavity where a background magnetic field induces an elliptic polarization. 
	The light bounces $\mathcal{O}(10^5)$ times between the mirrors $M_1$ and $M_2$, accumulating only the CP-even part of the ellipticity (proportional to $c-4b$, see Eq.~\eqref{eq:elliptEFTCPE}). 
	A polarizer $P_2$ is employed to extract the orthogonal part of the polarization, collected in the detector $D$. \emph{Right}: Modification of the PVLAS experiment. If the linear Fabry-Perot cavity is replaced by a ring cavity filled by the magnetic field only in its lower part, also the CP-odd part of the ellipticity (proportional to $d$) is accumulated over a round trip.} }
	\label{fig:PVLAScavity}
\end{figure}

The cancellation of the CP-odd contribution can be avoided by a modification of the optical path such that only part of it will be inside the magnetic field. 
Therefore we are motivated to consider a ring cavity instead of a linear cavity. 
If only the lower part of the ring cavity is permeated by the magnetic field, both the CP-even and the CP-odd parts of $\Psi$ and $\zeta$ would be accumulated, as the interaction with the magnetic field takes place exclusively for one propagation direction of the beam. 
A schematic design of the ring cavity proposal in comparison to the PVLAS design is presented in Fig.~\ref{fig:PVLAScavity}. 
Note that the essential difference between our ring cavity proposal and the PVLAS setup is the optical path.

Let us compare the PVLAS setup to the cavity ring proposal in more detail. 
In both, a magnetic field is slowly rotating in the plane perpendicular to the wave vector of the incoming light, with an angular frequency $\omega_B\ll\omega$, such that the approximation of static magnetic field holds.
A linearly polarized light (by the polarizer $P_1$) is fed into the cavity.  
While in the PVLAS setup it bounces between the mirrors $M_1$ and $M_2$, where the optical path is fully under the magnetic filed (left panel of Fig.~\ref{fig:PVLAScavity}), in the ring cavity setup the light will be bouncing between four mirrors, $M_{1,2,3,4}$, such that only part of the optical path is inside the magnetic field (right panel of Fig.~\ref{fig:PVLAScavity}). 
In order to increase the amplitude of the outgoing wave, a time-dependent ellipticity $\eta = \eta_0\cos(\omega_\eta t)$ is injected via a modulator, $G_\eta$.
In this way, the leading outgoing signal wave will be an interference between $\Psi$ and $\eta$, with a linear dependence on $\Psi$ rather than quadratic. 
The light detector ($D$) finally collects only the component of the polarization orthogonal to that of the incoming light, selected by $P_2$ (this component is nonzero thanks a nonvanishing ellipticity and rotation). 
Thus, the ratio between the incoming ($I_{\rm in}$) and outgoing ($I_{\rm out}$) wave intensities is~\cite{Millo:2008ug,DellaValle:2015xxa}
\begin{align}
	\label{eq:Aout}
	\frac{I_{\rm out}}{I_{\rm in}}
	\approx 
	\eta(t)^2
	+2\eta(t) \left[ N_e\Psi^e \leri{\alpha\leri{t}}+N_o\Psi^o\leri{\alpha\leri{t}}\right]\,,
\end{align}
where $\alpha(t)$ 
is the angle between the polarization vector $\mathbf{E}_i$ and  the magnetic field. 
The number of roundtrips inside the cavity is related to the cavity finesse $\mathcal{F}$ by $N\simeq \mathcal{F}/\pi$, where for PVLAS $\mathcal{F}\simeq7\cdot 10^5$. 
While the amplification factors for PVLAS are $N_e=2N+1$ and $N_o=1$, for the ring cavity they are $N_e=N_o=N+1$. As in PVLAS, to measure the rotation $\zeta$, one should insert a quarter-wave plate with one of its axes aligned along the initial polarization, such that the rotation is converted to an ellipticity, and interferes with $\eta$~\cite{Ejlli:2020yhk}. Thus, the signal becomes
\begin{equation}
\label{eq:Aout1}
\frac{I_{\rm out}}{I_{\rm in}}
\approx \eta(t)^2+2\eta(t)\left[N_e\zeta^e \leri{\alpha\leri{t}}+N_o\zeta^o\leri{\alpha\leri{t}}\right]\,.
\end{equation}

We note that although the polarization plane is not parallel to the mirrors in a ring cavity (unlike in the linear cavity), assuming the mirrors are properly aligned, the polarization vector $\mathbf{E}_i$ will not be shifted by the mirrors after a full round trip.\footnote{In addition, any misalignment can be accounted for by a control sample with the magnetic field turned off, for instance.} 
Therefore, in a ring cavity, the projected sensitivity for the CP-odd ellipticity and rotation could be similar as to their CP-even counterparts. Importantly, since the CP-even and CP-odd signals vary differently with $\alpha$, see Eqs.~\eqref{eq:scalar_birefringence}--\eqref{eq:elliptEFTCPE}, in the case of a measurement the two can be disentangled by a temporal analysis of the outgoing intensity in Eqs.~\eqref{eq:Aout} and~\eqref{eq:Aout1}, see~\cite{Millo:2008ug} for this analysis. A similar idea has been proposed in~\cite{Fan:2017sxk}, where instead the magnetic field does not rotate and the \ac{CPV} part of the photon self-interactions is selected by the choice of the (time-independent) angle $\alpha$.

\begin{figure}[t!]
	\centering
	\includegraphics[scale=0.59]{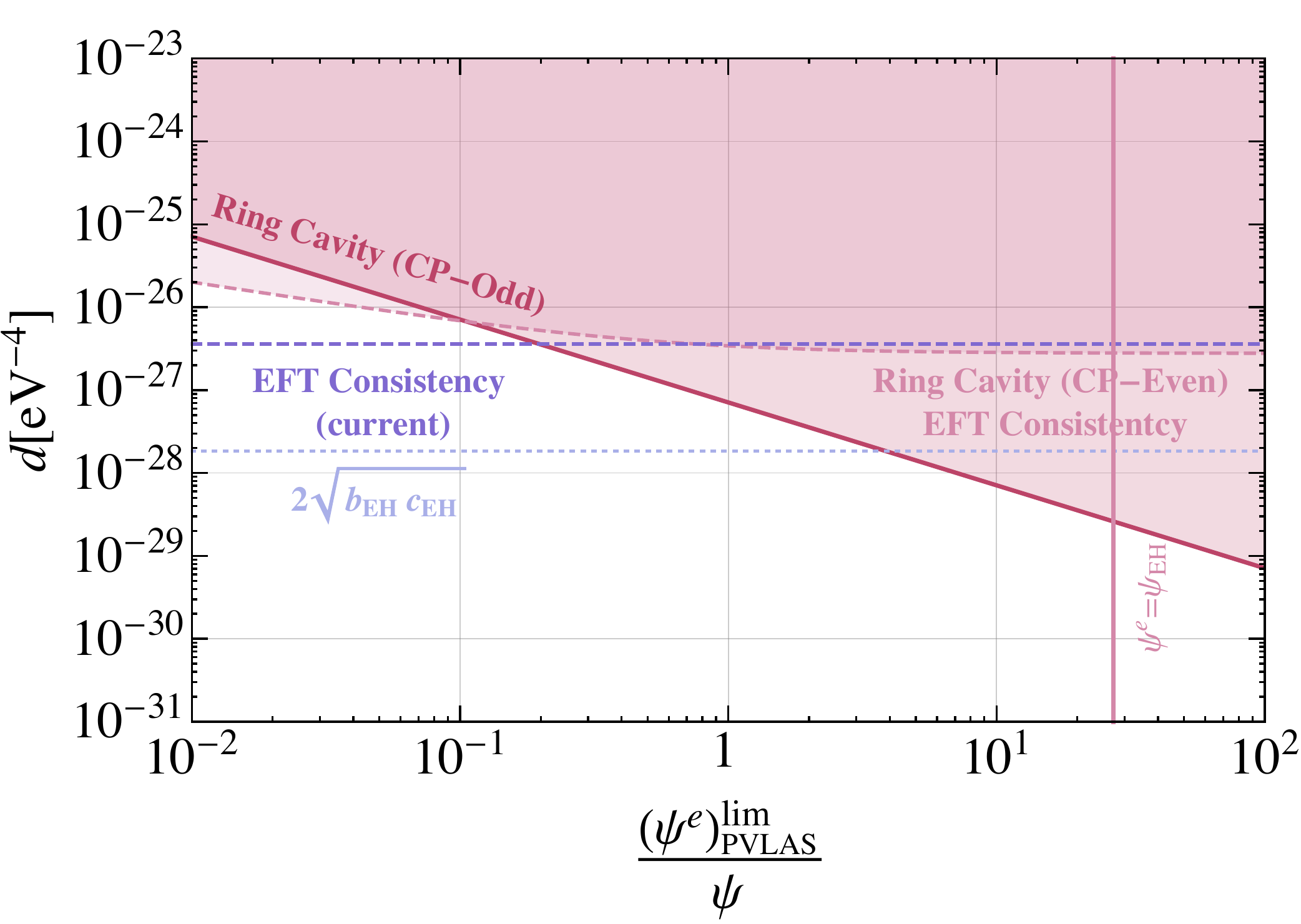}
	\caption{The expected bound on the coefficient $d$ of the effective operator $FFF\tilde{F}$ from vacuum birefringence in a ring cavity (solid red line), as a function of the experimental sensitivity to the ellipticity $\Psi$, normalized for convenience to the current sensitivity on this observable in the PVLAS experiment, $(\Psi^e)^{\rm lim}_\text{PVLAS}=
	8\cdot 10^{-12}$. 
	The upper horizontal dashed line is the \ac{EFT} consistency bound $d\leq2\sqrt{bc}$ with the current bounds on $b$ and $c$, as in Fig.~\ref{fig:SRFboundsd}. The lower horizontal dashed line corresponds to $d\leq2\sqrt{b_\text{EH}c_\text{EH}}$, which would be the consistency bound on the \ac{CPV} coefficient assuming the \ac{EH} term would be measured. The pink dashed line corresponds to the prospective consistency bound from the improvement in the bound on $c$ considering only the measurement of the \ac{CPC} signal (using the same cavity). 
	The vertical line corresponds to the sensitivity to $\Psi^e$ at which the~\ac{EH} contribution should be observed (analyzing the CP-even part of the signal). In particular, for $\Psi^e>\Psi_{\rm EH}$, the pink dashed line is plotted assuming that \ac{EH} background can be fully removed in the measurement of the \ac{CPC} part of the signal.
	} 
	\label{EFTfigRing}
\end{figure}

By inverting Eq.~\eqref{eq:elliptEFTCPE}, we can express the reach of the ring cavity in the measurement of $d$ in terms of the minimum measurable ellipticity $\leri{\Psi^o}^\text{lim}$ (acquired over the full set of round trips) as
 \begin{align}\label{eq:dlim}
	 |d^\text{lim}|
=	 \frac{\pi}{ B^2\omega L \mathcal{F}} \leri{\Psi^o}^\text{lim}\, .
 \end{align}
As a reference, the PVLAS sensitivity for the CP-even part of the ellipticity is $\leri{\Psi^e}^{\text{lim}}_{\rm PVLAS}=2 \sigma_{\Psi }^\text{PVLAS}=8\cdot 10^{-12}$, which is the latest result, reported in~\cite{Ejlli:2020yhk}, obtained over a measurement time of $5\times 10^6$~s.
This value, using $B_\text{PVLAS}=2.5\,$T, $L_\text{PVLAS}=0.82\,$m and $\omega_{\rm PVLAS}=1.2\,$eV, provides the bound on $c-4b$ in Eq.~\eqref{eq:bcPVLAS}.

In Fig.~\ref{EFTfigRing} we present the potential reach of the ring cavity for the measurement of $d$. 
We show the bound as a function of the minimum measurable ellipticity, normalized for reference to the current bound on this observable by the PVLAS experiment (quoted above). In the plot we assume that the same magnetic field, finesse and cavity length will be employed. If the dominant noise is independent of the finesse, the magnetic field and the length (see~\cite{Ejlli:2020yhk} for a discussion on the validity of this assumption), the relative bound on $d$ compared to the PVLAS bound on $c-4b$ scales as 
\begin{align}
	d^\text{ring} 
	\propto 
	\frac{B^2_\text{PVLAS}}{B^2}
	\frac{2\mathcal{F}_\text{PVLAS}}{\mathcal{F}}
	\frac{L_\text{PVLAS}} {L} (c-4b)^\text{PVLAS}\,, 
\end{align}
where the factor of $2$ comes from the fact that the light traveling in the \ac{FP} cavity crosses the magnetic field twice at each round trip. 
In this case increasing either $B$ or $L$ could greatly improve the bound.

Since the ring cavity would be sensitive also to the combination $c-4b$, the bound on this quantity could get stronger, and correspondingly that on $c$ (obtained as in Eq.~\eqref{eq:cComb}, assuming the same bound on $b$ of Eq.~\eqref{eq:bLambShift}). 
Therefore, we also present the prospective \ac{EFT} consistency condition of Eq.~\eqref{eq:EFTcon} with this updated $c$. 
As the bound is presented as a function of the improved sensitivity, in Fig.~\ref{EFTfigRing} we mark the value of $\leri{\Psi^e}^{\text{lim}}_{\rm PVLAS}/\Psi$ at which the \ac{EH} contribution to the photon self-interactions will be observed, and assume it will be removed completely when deriving the bound. We observe that the left hand side of Eq.~\eqref{eq:EFTcon} is proportional only to $\sqrt{c}$. 
Thus, using a ring cavity, an increase in the sensitivity for vacuum birefringence would improve both the bounds on $c$ and on $d$, but (assuming the same bound on $b$ of Eq.~\eqref{eq:bLambShift}) still probing values of $c$ and $d$ that are compatible with the unitarity and causality of the UV theory. 

A similar discussion applies to the ellipticity and rotation in Eqs.~\eqref{eq:elliptphi} and~\eqref{eq:rotphi} generated by a light scalar, both of which, as mentioned, can be split into a CP-even and CP-odd part. In a ring cavity the bound on $\sqrt{g\tilde{g}}$ is obtained both from the measurement of the ellipticity and from the measurement of the rotation, and we show the strongest of the two, \ie
\begin{align}\label{eq:ggtlimmmin}
|g\tilde{g}|^\text{lim}
=&	\frac{\pi}{B^2 \mathcal{F}}
	\text{Min}\leri{
	 \frac{\abs{\leri{\Psi^o}^\text{lim}}}{2\left(1-\frac{\sin x}{x}\right)\frac{\omega L}{4m_\phi^2}},
	\frac{\abs{\leri{\zeta^o}^\text{lim} }}{2\sin^2{\frac{x}{2}}\frac{\omega^2}{m_\phi^4}}}\,,
\end{align}
in Fig.~\ref{fig:SRFboundsggt} (red line). In plotting the bound we assumed that the ring cavity will be able to get to the same sensitivity as PVLAS, \ie~$\leri{\Psi^o}^\text{lim}=\leri{\Psi^e}^{\text{lim}}_{\rm PVLAS}/2= \sigma^\text{PVLAS}_{ \Psi}$ and $\leri{\zeta^o}^\text{lim}=\leri{\zeta^e}^\text{lim}_\text{PVLAS}/2= \sigma^\text{PVLAS}_{ \zeta}$, with the same magnetic field and cavity length. Notice that bound on $\sqrt{g\tilde{g}}$ from the rotation dominates at small masses and becomes $m_\phi$-independent for $m_\phi/\omega\ll1$ (see Fig.~\ref{fig:SRFboundsggt}), while the one from the ellipticity dominates at higher masses, for which $|g\tilde{g}|\,^\text{lim}\propto m_\phi^2$ (see Eq.~\eqref{eq:ggtlimmmin}). As expected, these limits are of the same order as other laboratory bounds and not yet competitive with other current bounds. For the fermionic dipole operators, our bound (assuming $\Psi\approx  \leri{\Psi^e}^{\text{lim}}_{\rm PVLAS}$ and similar finesse) yields $D\,, \tilde{D}\lesssim 10^{-7} \text{eV}^{-1}$.

\section{Conclusions}
\label{sec:conclusions}
\acresetall
In this paper, we considered the possibility that the photon is subject to \ac{CPV} self-interactions, encoded in the low-energy effective operator $F_{\mu\nu}F^{\mu\nu}F_{\rho\sigma}\tilde{F}^{\rho\sigma}$,  with ($\tilde F_{\mu\nu}$) $F_{\mu\nu}$ being the (dual) electromagnetic field strength,
 which are highly suppressed in the \ac{SM} but could receive contributions from new physics. 

We discussed two simple experimental approaches to isolate such interactions at energies below the electron mass. 
In particular, we estimated the corresponding sensitivities of two table-top cavity experiments to the above non-linear operator -- one using a \ac{SRF} cavity and one using a ring \ac{FP} cavity, see Figs.~\ref{fig:SRFboundsd} and~\ref{fig:PVLAScavity}. These are expected to give the first direct limit on CP violation in non-linear \ac{EM} in vacuum.

In passing, we also qualitatively estimated the indirect bounds on the above \ac{CPV} operator at energies above the electron mass. We found that at these energies the indirect bounds from the experimental limits on the electric and magnetic dipole moments of the electron are stronger, and therefore provide more stringent constraints on possible generic new physics if heavier than the electron.

In addition to considering an \ac{EFT} approach towards the \ac{CPV} effects discussed above, we derived the corresponding bounds for the case where \ac{CPV} photon interactions are mediated by light scalars and fermions. 
We found that the constraints on the \ac{CPV} combination of the couplings of a scalar to photons are not competitive with fifth-force searches and searches for violation of the equivalence principle. 
We further found that our experimental setup would be able to reach world-record sensitivity to the presence of new fermions with electric and magnetic dipole moments, provided their mass lies between few hundreds keV to 10\,MeV (for masses smaller than few hundreds keV, astrophysical bounds become stronger).  
However, one can indirectly set a stronger bound on the corresponding dipole interactions (at least for fermions heavier than the electron) using the constraints associated with the electric dipole moment of the electron. 

Finally, we note that in the two \ac{SRF}-cavities setup proposed in Ref.~\cite{Gao:2020anb,Salnikov:2020urr}, it might be possible to constrain the \ac{CPV} part of the non-linear \ac{EM} interactions, in a manner similar to this work.

\section*{Acknowledgments}
We thank Nitzan Ackerman and Roee Ozeri for useful discussions. We are grateful to Roni Harnik and especially Yoni Kahn for comments on the draft. 
The work of GP is supported by grants from BSF-NSF, the Friedrich Wilhelm Bessel research and Segre awards, GIF, ISF, Minerva and Yeda-Sela-SABRA-WRC. 
YS is a Taub fellow (supported by the Taub Family Foundation) and is supported by grants from NSF-BSF, ISF and the Azrieli foundation. IS~is supported by a fellowship from the Ariane de Rothschild Women Doctoral Program.

\appendix

\section{Cavity Modes}\label{sec:cavitymodes}

\begin{table}[t!]
	\centering
	\begin{tabular}{|c|c|c|c|c|c|}
		\hline
		\textbf{$\omega_1$}&\textbf{$\omega_2$}&\textbf{$\omega_s$}&
		\textbf{$h/R$}&\textbf{$K_0$}&\textbf{$K_\infty$}\\ \hline
	$\text{TE}_{011}$& $\text{TM}_{040}$ & $\text{TM}_{060}$ & 0.39643 & 0.091 & 0.18 \\\hline
	$\text{TE}_{011}$& $\text{TM}_{050}$& $\text{TM}_{080}$ & 0.396798 & 0.45 & 0.13 \\\hline
		$\text{TE}_{021}$&$\text{TM}_{050}$&$\text{TM}_{060}$ & 0.165757 & 0.26 & 0.25 \\\hline
	$\text{TE}_{021}$&$\text{TM}_{050}$&$\text{TM}_{070}$ & 0.310411 & 0.31 & 0.30 \\\hline
		$\text{TE}_{023}$&$\text{TM}_{040}$&$\text{TM}_{050}$ & 0.930618 & 0.13 & 0.15 \\\hline
		$\text{TE}_{031}$&$\text{TM}_{060}$&$\text{TM}_{080}$ & 0.263649 & 0.25 & 0.39 \\\hline
		$\text{TE}_{033}$&$\text{TM}_{050}$&$\text{TM}_{060}$ & 0.790656 & 0.24 & 0.19 \\\hline
		$\text{TE}_{033}$&$\text{TM}_{060}$&$\text{TM}_{070}$ & 0.431223 & 0.20 & 0.11 \\\hline
		$\text{TE}_{033}$&$\text{TM}_{060}$&$\text{TM}_{080}$& 0.790948 & 0.82 & 0.13 \\\hline
		$\text{TE}_{033}$&$\text{TM}_{070}$&$\text{TM}_{080}$ & 0.315526 & 0.783 & 0.10\\\hline
		$\text{TE}_{035}$&$\text{TM}_{050}$&$\text{TM}_{060}$ & 1.31776 & 0.94 & 0.11 \\\hline
\end{tabular}
\caption{TE/TM/TM mode combinations with $K_\infty, K_0 \gtrsim 0.1$ for \ac{CPV}  interactions.}
\label{modes-TM}
\end{table}

\begin{table}[t!]
	\centering
	\begin{tabular}{|c|c|c|c|c|c|}
		\hline
		\textbf{$\omega_1$}&\textbf{$\omega_2$}&\textbf{$\omega_s$}&\textbf{$h/R$}&\textbf{$K_0$}&\textbf{$K_\infty$}\\ \hline
	$\text{TM}_{020}$&$\text{TE}_{023}$& $\text{TE}_{031}$ & 1.31175 & 0.15& 0.35 \\\hline
	$\text{TM}_{030}$&$\text{TE}_{012}$&$\text{TE}_{033}$ & 0.162995 & 0.12& 0.12 \\\hline
	$\text{TM}_{030}$&$\text{TE}_{022}$&$\text{TE}_{023} $& 0.251037 & 0.20& 0.26 \\\hline
	$\text{TM}_{030}$&$\text{TE}_{022}$&$\text{TE}_{033}$ & 0.20255 & 0.11& 0.24 \\\hline
	$\text{TM}_{030}$&$\text{TE}_{023}$&$\text{TE}_{025}$ & 0.221669 & 0.12 & 0.13 \\\hline
	$\text{TM}_{030}$&$\text{TE}_{023}$&$\text{TE}_{031}$ & 0.731319 & 0.13& 0.99 \\\hline
	$\text{TM}_{030}$&$\text{TE}_{023}$&$\text{TE}_{033}$ & 0.564125 & 0.31& 0.32 \\\hline
	$\text{TM}_{030}$&$\text{TE}_{023}$&$\text{TE}_{035}$ & 0.198568 & 0.51 & 0.12 \\\hline
	$\text{TM}_{030}$&$\text{TE}_{032}$&$\text{TE}_{025}$ & 0.660409 & 0.14& 0.22 \\\hline
	$\text{TM}_{030}$&$\text{TE}_{032}$&$\text{TE}_{035}$& 1.07868 & 0.24& 0.37 \\\hline
\end{tabular}
\caption{TM/TE/TE mode combinations with $K_\infty, K_0 \gtrsim 0.1$ for \ac{CPV}  interactions. }
\label{modes-TE}
\end{table}

We list a few mode choices achieving sizable form factors $K_0$ and/or $K_\infty$ for a right cylindrical \ac{SRF} cavity. Table~\ref{modes-TM} corresponds to $\text{TE}_{0p_1 q_1}/\text{TM}_{0p_2 0}/\text{TM}_{0p_s0}$ mode combinations, and Table~\ref{modes-TE} corresponds to $\text{TM}_{0p_10}/\text{TE}_{0p_2 q_2}/\text{TE}_{0p_s q_s}$ combinations.

\section{Derivation of The Refractive Indices}
\label{sec:ellipt}
In this Appendix we derive the refractive indices and the eigen-axes of propagation for a linearly polarized light beam (with electric field $\mathbf{E}$ and magnetic field $\mathbf{B}$) traveling through a region permeated by a constant and homogeneous background magnetic field $\mathbf{B}_0$ (orthogonal to the beam), in the presence of the photon self-interactions with the Lagrangian in Eq.~\eqref{eq:LEFT}. We will assume that the light beam field is negligible with respect to the background field, \ie~that $\mathbf{E}\ll \mathbf{B}_0$ and $\mathbf{B}\ll \mathbf{B}_0$. The derivation below mostly follows Refs.~\cite{Millo:2008ug,PhysRevA.63.012107}.

The Euler-Lagrange equations of the Lagrangian in Eq.~\eqref{eq:LEFT} 
can be equivalently written in terms of the electric and magnetic fields $\mathcal{E}_i\equiv F_{0i}$ and $\mathcal{B}_i\equiv -\frac12 \epsilon_{ijk}F^{jk}$ as
\begin{align}
	\label{eq:maxw1}
	\div{ \boldsymbol{\mathcal{B}}}&=0\,, & \div{ \mathbf{D}}&=0\,,\\
	\curl{ \boldsymbol{\mathcal{E}}}&=-\partial_t \boldsymbol{\mathcal{B}}\,, & \curl{ \mathbf{H}}&=\partial_t \mathbf{D}\,,\label{eq:maxw2}
\end{align}
where $\mathbf{D}$ and $\mathbf{H}$ 
are defined by
\begin{align}
	\label{eq:DE}
	\mathbf{D}\equiv\pdv{\mathcal{L}_{\rm EFT}}{ \boldsymbol{\mathcal{E}}} \, , \qquad\qquad
	\mathbf{H}\equiv-\pdv{\mathcal{L}_{\rm EFT}}{ \boldsymbol{\mathcal{B}}}\,.
\end{align}
In the setup under consideration, the fields are $\boldsymbol{\mathcal{E}}=\mathbf{E}$ and $\boldsymbol{\mathcal{B}}=\mathbf{B}+\mathbf{B}_0$. From Eqs.~\eqref{eq:LEFT} and~\eqref{eq:DE} it follows that
\begin{align}
	\label{eq:D1}
	\mathbf{D}
=&	- \mathbf{E}+4bB_0^2 \mathbf{E}
	+\frac{d}{2}B_0^2 \mathbf{B}
	-2c \mathbf{E}\cdot \mathbf{B}_0 \mathbf{B}_0
	+2d \mathbf{B}\cdot \mathbf{B}_0 \mathbf{B}_0\,,\\
	\mathbf{H}=&- \mathbf{B}+4bB_0^2 \mathbf{B}
	-\frac{d}{2}B_0^2 \mathbf{E}
	+8b \mathbf{B}\cdot \mathbf{B}_0 \mathbf{B}_0
	-2d \mathbf{E}\cdot \mathbf{B}_0 \mathbf{B}_0\,,\label{eq:H1}
\end{align}
where terms with higher powers of $\mathbf{E}$ or $\mathbf{B}$ were omitted since they are negligible with respect to the last three terms of Eqs.~\eqref{eq:D1} and~\eqref{eq:H1}, given the assumption $\mathbf{E}\ll \mathbf{B}_0$ and $\mathbf{B}\ll \mathbf{B}_0$. Notice that Eqs.~\eqref{eq:D1} and~\eqref{eq:H1} have the form
\begin{align}
	D_i&=\epsilon_{ij}E_j+\psi^\text{DB}_{ij}B_j\,,\\
	H_i&=\mu_{ij}B_j+\psi^\text{HE}_{ij}E_j\,,
\end{align}
where $\epsilon_{ij}$, $\mu_{ij}$, $\psi^\text{DB}_{ij}$ and $\psi^\text{HE}_{ij}$ are constant matrices that depend on $\mathbf{B}_0$ and on the photon self-interaction coefficients ($\epsilon_{ij}$ and $\mu_{ij}$ are the equivalent of the electric permittivity and magnetization tensors). These matrices can be read off of Eqs.~\eqref{eq:D1} and~\eqref{eq:H1}.

To solve Eqs.~\eqref{eq:maxw1} and~\eqref{eq:maxw2} together with Eqs.~\eqref{eq:D1} and~\eqref{eq:H1} for the light beam, we consider the ansatz
\begin{align}
	\label{eq:EEtilde}
	\mathbf{E}
=	\tilde{\mathbf{E}}\exp[i(\omega t-\mathbf{k}\cdot\mathbf{x})]\,,
\end{align}
 which corresponds to a plane wave with momentum $\mathbf{k}$ and frequency $\omega$.  Substituting Eq.~\eqref{eq:EEtilde} and using $\partial_t \mathbf{B}_0=0$, the first of Eq.~\eqref{eq:maxw2} becomes $\mathbf{k}\times\tilde{\mathbf{E}}=\omega\tilde{\mathbf{B}}$, while the second is $\mathbf{k}\times(\mu\tilde{\mathbf{B}}+\psi^{\rm HE}\tilde{\mathbf{E}})=-\omega(\epsilon\tilde{\mathbf{E}}+\psi^{\rm DB}\tilde{\mathbf{B}})$. These two equations can be combined together to eliminate $\tilde{\mathbf{B}}$ to give, in matrix form,
\begin{align}
	n^2\hat{k}\times\leri{\mu\hat{k}\times\tilde{\mathbf{E}}}
	+n\hat{k}\times\leri{\psi^\text{HE}\tilde{\mathbf{E}}}
	+n\psi^\text{DB}\hat{k}\times\tilde{\mathbf{E}}+\epsilon \tilde{\mathbf{E}} \, 
	=0\,,
\end{align}
where we defined the refractive index $n$ by $n\equiv k/\omega$, see also Eq.~\eqref{eq:Etx} in the main text. From the explicit form of the matrices $\epsilon_{ij}$, $\mu_{ij}$, $\psi^\text{DB}_{ij}$ and $\psi^\text{HE}_{ij}$ we get 
\begin{align}
	\label{eq:dispB0}
	n^2&\hat{k}\times\leri{\leri{4bB_0^2-1}\hat{k}\times \tilde{\mathbf{E}}
	+8b\hat{k}\times \tilde{\mathbf{E}}\cdot \mathbf{B}_0 \mathbf{B}_0}
	+4bB_0^2 \tilde{\mathbf{E}}-2c \tilde{\mathbf{E}}\cdot \mathbf{B}_0 \mathbf{B}_0\\\nonumber
&	-2dn \tilde{\mathbf{E}}\cdot \mathbf{B}_0\hat{k}\times \mathbf{B}_0
	+2dn\leri{\hat{k}\times \tilde{\mathbf{E}}\cdot \mathbf{B}_0 \mathbf{B}_0}= \tilde{\mathbf{E}}\,.
\end{align}
We assume the light beam to travel perpendicularly to $\mathbf{B}_0$, \ie~$\mathbf{B}_0\cdot\hat{\mathbf{k}}=0$, and that the beam's momentum while propagating in the magnetic field is $\hat{\mathbf{k}}=\pm\hat{\mathbf{k}}_i$, where $\hat{\mathbf{k}}_i$ is the propagation direction of the initial beam. Therefore, $\set{\mathbf{\hat{B}}_0,\hat{\mathbf{k}}_i\times\mathbf{\hat{B}}_0}$ forms a constant orthonormal basis spanning the polarization plane of the beam, where $\mathbf{E}$ lies. In this convenient basis, Eq.~\eqref{eq:dispB0} can be brought into the matrix form
\begin{align}
	B_0^2
	\begin{pmatrix}
		4b\leri{1-n^2}-2c& -2  d n \leri{\hat{\mathbf{k}}\cdot\hat{\mathbf{k}}_i} \\
		-2  d n \leri{\hat{\mathbf{k}}\cdot\hat{\mathbf{k}}_i} & 4b\leri{1-n^2}-8b
	\end{pmatrix}
	\begin{pmatrix} 
		\tilde{\mathbf{E}}\cdot\mathbf{\hat{B}}_0 \\ 
		\tilde{\mathbf{E}}\cdot\leri{\hat{\mathbf{k}}_i\times\mathbf{\hat{B}}_0} 
	\end{pmatrix}
= 	\leri{1-n^2}\begin{pmatrix} 
		\tilde{\mathbf{E}}\cdot\mathbf{\hat{B}}_0 \\ 
		\tilde{\mathbf{E}}\cdot\leri{\hat{\mathbf{k}}_i\times\mathbf{\hat{B}}_0} 
	\end{pmatrix}\,.
\end{align}
In the limit $b B_0^2,cB_0^2,dB_0^2\ll 1$, and expanding at leading order in $n-1$, the previous equation is simplified as
\begin{align}
	\label{eq:nMatrixEFT}
	\begin{pmatrix}
		1+cB_0^2&  B_0^2 d \leri{\hat{\mathbf{k}} \cdot\hat{\mathbf{k}}_i} \\
		B_0^2 d \leri{\hat{\mathbf{k}}\cdot\hat{\mathbf{k}}_i} & 1+4bB_0^2
	\end{pmatrix}
	\begin{pmatrix} 
		\tilde{\mathbf{E}}\cdot\mathbf{\hat{B}}_0 \\ 
		\tilde{\mathbf{E}}\cdot\leri{\hat{\mathbf{k}}_i\times\mathbf{\hat{B}}_0} 
	\end{pmatrix}
= 	n\begin{pmatrix} 
		\tilde{\mathbf{E}}\cdot\mathbf{\hat{B}}_0 \\ 
		\tilde{\mathbf{E}}\cdot\leri{\hat{\mathbf{k}}_i\times\mathbf{\hat{B}}_0} 
	\end{pmatrix}\,.
\end{align}
The eigenvectors of the matrix in Eq.~\eqref{eq:nMatrixEFT} are the propagation eigenstates, denoted in the main text as $\hat{v}_1$ and $\hat{v}_2$, in the basis $\set{\mathbf{\hat{B}}_0,\hat{\mathbf{k}}_i\times\mathbf{\hat{B}}_0}$. Any vector proportional to one of the propagation eigenstate solves Eq.~\eqref{eq:nMatrixEFT} (and therefore the the initial equations of motion with ansatz in Eq.~\eqref{eq:EEtilde}) provided the refractive index coincides with the eigenvalues of the matrix above, which are $n_{1,2}$ in Eq.~\eqref{eq:n12}.

As mentioned in the main text, for $d=0$ the eigenvectors $\set{\hat{v}_1, \hat{v_2}}$ coincide with $\set{\mathbf{\hat{B}}_0,\hat{\mathbf{k}}_i\times\mathbf{\hat{B}}_0}$ and are otherwise rotated with respect to $\set{\mathbf{\hat{B}}_0,\hat{\mathbf{k}}_i\times\mathbf{\hat{B}}_0}$ by the angle $\alpha_{\cancel{\rm CP}}$ in Eq.~\eqref{eq:alphacpv}, which positive direction is $\mathbf{\hat{B}}_0\times\hat{\mathbf{k}}_i\times\mathbf{\hat{B}}_0=\hat{\mathbf{k}}_i$. Moreover, flipping the sign of $\mathbf{k}$ is equivalent to $d\to-d$ as Eq.~\eqref{eq:nMatrixEFT} remains invariant, and this corresponds to changing the sign of $\alpha_{\cancel{\rm CP}}$, given Eq.~\eqref{eq:alphacpv}.

\bibliographystyle{JHEP}
\bibliography{cavities}


\end{document}